\documentclass[10pt]{article}
\usepackage{fancyhdr}
\usepackage{extramarks}
\usepackage{amsmath}
\usepackage{amsthm}
\usepackage{amsfonts}
\usepackage{siunitx}
\usepackage{tikz}
\usepackage[plain]{algorithm}
\usepackage{algpseudocode}
\usepackage{multirow}
\usepackage{booktabs}
\usepackage{graphicx}
\usepackage{subfigure}
\usepackage{xcolor}

\usepackage{longtable}
 \usepackage{color}
\usepackage{amsmath,bm}
\usepackage{booktabs}
\usepackage{mathtools}
\usepackage{amssymb}
\usepackage{caption}
\usepackage{capt-of}
\usepackage{mciteplus}
\usepackage{cite}

\usepackage{mathrsfs}
\usepackage[title,titletoc,toc]{appendix}
\usepackage{xr}
\usepackage{parskip}
\usepackage{soul}
\usepackage{textcomp}
\usepackage[colaction]{multicol}
\usepackage[switch]{lineno}
\usepackage{lipsum}
\usepackage{etoolbox}
\usepackage{longtable}
\usepackage{array}
\usepackage{tablefootnote}
\usepackage{ragged2e}
\newcolumntype{C}[1]{>{\centering\arraybackslash}p{#1}}
\usetikzlibrary{automata,positioning}
\usepackage{adjustbox}
\usepackage{makecell}
\usepackage{pdflscape}

\usepackage{booktabs}
\usepackage{array}
\usepackage{caption}
\usepackage{tabularx}
\usepackage{booktabs}
\DeclareCaptionFormat{custom_format}{%
  \textbf{#1~#2}%
  \vspace{0.3pt}
  \par\rule{\textwidth}{0.5pt}\par%
  \vspace{0.5pt}
  #3%
}

\usepackage{setspace}

\usepackage{booktabs}
\usepackage{url} 

\usepackage[font=small, labelfont=bf]{caption} 
\usepackage{multicol}
\usepackage[utf8]{inputenc}
\usepackage[T1]{fontenc}
\usepackage{textgreek}
\usepackage{xcolor}
\usepackage{url}
\usepackage[colorlinks=true,linkcolor=black,anchorcolor=black,citecolor=black,urlcolor=blue]{hyperref}

\usetikzlibrary{automata,positioning}

\usepackage{graphicx} 

\begin{document}
\title{Meta-analysis and Topological Perturbation in Interactomic Network for Anti-opioid Addiction Drug Repurposing} 
\author{Chunhuan Zhang$^{1}$, Sean Cottrell$^{2}$, Benjamin Jones$^2$, Yueying Zhu$^{1}$, Huahai Qiu$^{1}$, \\Bengong Zhang$^{1}$, Tianshou Zhou$^{3}$, and Jian Jiang$^{1,2}$\footnote{Corresponding author.  Email: jjiang@wtu.edu.cn} \\
$^{1}$Research Center of Nonlinear Science, School of
Mathematics and Statistics, \\Wuhan Textile University, Wuhan, 430200, P R. China \\
$^{2}$Department of Mathematics, Michigan State University, East Lansing, \\Michigan 48824, USA\\
$^{3}$Key Laboratory of Computational Mathematics, Guangdong Province, and School of \\Mathematics, Sun Yat-sen University, Guangzhou, 510006, P R. China }


\date{\today} 

\maketitle

\abstract{The ongoing opioid crisis highlights the urgent need for novel therapeutic strategies that can be rapidly deployed. This study presents a novel approach to identify potential repurposable drugs for the treatment of opioid addiction, aiming to bridge the gap between transcriptomic data analysis and drug discovery. Specifically, we perform a meta-analysis of seven transcriptomic datasets related to opioid addiction by differential gene expression (DGE) analysis, and propose a novel multiscale topological differentiation to identify key genes from a protein-protein interaction (PPI) network derived from DEGs. This method uses persistent Laplacians to accurately single out important nodes within the PPI network through a multiscale manner to ensure high reliability. Subsequent functional validation by pathway enrichment and rigorous data curation yield 1,865 high-confidence targets implicated in opioid addiction, which are cross-referenced with DrugBank to compile a repurposing candidate list. To evaluate drug-target interactions, we construct predictive models utilizing two natural language processing-derived molecular embeddings and a conventional molecular fingerprint. Based on these models, we prioritize compounds with favorable binding affinity profiles, and select candidates that are further assessed through molecular docking simulations to elucidate their receptor-level interactions. Additionally, pharmacokinetic and toxicological evaluations are performed via ADMET (absorption, distribution, metabolism, excretion, and toxicity) profiling, providing a multidimensional assessment of druggability and safety. This study offers a generalizable approach for drug repurposing in other complex diseases beyond opioid addiction.

}

\maketitle

 {\setcounter{tocdepth}{4} \tableofcontents}
\newpage
\section{Introduction}
The opioid addiction crisis poses a significant global public health challenge, particularly in the United States, where addiction rates and overdose-related deaths have escalated in recent years. At the heart of this crisis lies opioid use disorder (OUD), a chronic relapsing condition characterized by compulsive opioid use, loss of control over drug intake, and the emergence of negative emotional states during withdrawal. OUD is associated with profound medical, psychological, and socioeconomic consequences, including increased risk of infectious diseases, mental health disorders, and premature death. Despite the availability of some pharmacological treatments such as methadone and buprenorphine, their efficacy is limited by issues such as dependence, side effects, and poor patient adherence \cite{shulman2019buprenorphine, suarez2022buprenorphine}. These challenges underscore the pressing need for innovative therapeutic strategies to identify novel drug targets and repurpose existing compounds for opioid-addiction treatment. 

Recent advances in high-throughput transcriptomic technologies provide many opportunities to uncover molecular mechanisms underlying complex disorders like opioid addiction, offering insights into dysregulated genes and pathways that may serve as therapeutic entry points.  Traditional approaches to transcriptomic analysis primarily rely on differential gene expression (DEG) analysis, which is a crucial tool in molecular biology and genetics, showing promise in identifying new targets for drug addiction treatment, and aiding in discovering novel therapeutic agents \cite{wang2025darg}. By analyzing transcriptomic changes across diverse biological states, including those induced by substance exposure, DEG analysis facilitates the identification of genes whose expression levels vary significantly under differing conditions or among distinct sample populations. When gene expression patterns are contrasted between drug-exposed and control groups, researchers can gain insights into the molecular mechanisms that may underlie addictive behaviors or substance response, thereby informing early-stage target discovery. For example, Nestler et al. characterized the cell-type-specific restructuring of the nucleus accunmbens transcriptional landscape after opioid exposure integrating with multiscale embedded gene co-expression network analysis to uncover the molecular mechanisms governing substance use disorder pathology \cite{browne2025cocaine}. Carter et al. leveraged human prefrontal cortex RNAseq data from four independent opioid overdose death studies and conducted a transcriptome-wide DGE meta-analysis, which identified 335 significant differentially expressed genes from 20098 genes \cite{carter2024identifying}.

Despite the valuable insights offered by DEG-based studies, most rely on conventional algorithms to pinpoint key genes within protein-protein interaction (PPI) networks derived from DEGs. These algorithms typically emphasize topological connectivity, i.e. the existence of links between nodes, while often overlooking quantitative measures such as interaction confidence or binding strength. As a result, critical biological information may be lost. Additionally, these approaches are generally confined to analyzing low-dimensional relational structures, which limits their ability to capture the complex, high-dimensional architecture inherent in biological networks. Notably, although some of these methods succeed in identifying putative key genes, they frequently fall short of employing quantitative frameworks to support downstream drug discovery and repurposing efforts based on those targets.

To address these limitations, our study introduces a novel multiscale topological differentiation (MTD) framework, leveraging persistent Laplacians (also called persistent spectral graph) to extract topological signatures from PPI networks constructed from DEGs. Persistent Laplacian is a new algebraic topology tool introduced by Wang et al.\cite{wang2020persistent}  for capturing the molecular structure complexity and high dimensionality, which has been successfully employed in various applications, including drug addiction analysis \cite{zhu2023tidal}, protein-ligand binding affinity prediction \cite{meng2021persistent}, machine learning (ML)-assisted protein engineering \cite{qiu2023persistent}, predicting emerging SARS-CoV-2 variants \cite{chen2022persistent}, and so on. The MTD method captures the intrinsic geometry of molecular interactions across scales and enables the identification of structurally central genes that may play pivotal roles in addiction pathology. By integrating topological data analysis into the gene prioritization process, we expand the analytical landscape beyond conventional DEG-based strategies, offering a more reliable approach to target discovery.

Additionally, most existing transcriptomic studies rely heavily on single-dataset analyses from the Gene Expression Omnibus (GEO) database, which are inherently limited by small sample sizes, study-specific biases, and technical variability. Such approaches often yield inconsistent results, with candidate genes failing to replicate across cohorts, thereby impeding the identification of robust, generalizable molecular targets. Du et al. identified key genes with opioid and cocaine addiction from single dataset and three pivotal molecular targets, mTOR, mGluR5, and NMDAR, for drug repurposing from DrugBank \cite{du2024multiscale}. Here, we introduce a meta-analysis approach that aggregates multiple transcriptomic datasets from the GEO database. By harmonizing and analyzing data across diverse cohorts, this strategy enhances statistical power, mitigates batch effects, and captures the heterogeneity inherent in opioid addiction (e.g., variations in opioid type, exposure duration, and participant demographics). This multi-dataset integration ensures that identified genes are consistently dysregulated across contexts, increasing confidence in their relevance as biological markers and improving the generalizability of findings.

Drug repurposing, re-evaluating approved or investigational drugs for new therapeutic uses beyond their original indications, has demonstrated notable success across various disease contexts, offering a promising strategy to reduce both the cost and duration of drug development \cite{chen2024machine,jiang2025proteomic}. As biological datasets continue to expand rapidly, ML has emerged as a powerful tool in modern drug discovery pipelines. By applying nonlinear regression techniques to existing datasets, ML models are capable of uncovering hidden patterns that inform therapeutic relevance. Given the inherent complexity and high dimensionality of biomedical data, ML-based virtual screening often outperforms traditional physics-driven methods such as molecular docking and molecular dynamics (MD) simulations, particularly in accelerating the evaluation of vast chemical spaces. For example, Feng et al. applied ML algorithms to screen DrugBank compounds for potential interactions with MOR, KOR, and DOR, subsequently conducting detailed assessments of binding conformations and drug-likeness for candidates predicted to exhibit strong target affinity \cite{feng2023machine}.

\begin{figure*}[htp]
    \centering
    \includegraphics[width=14cm]{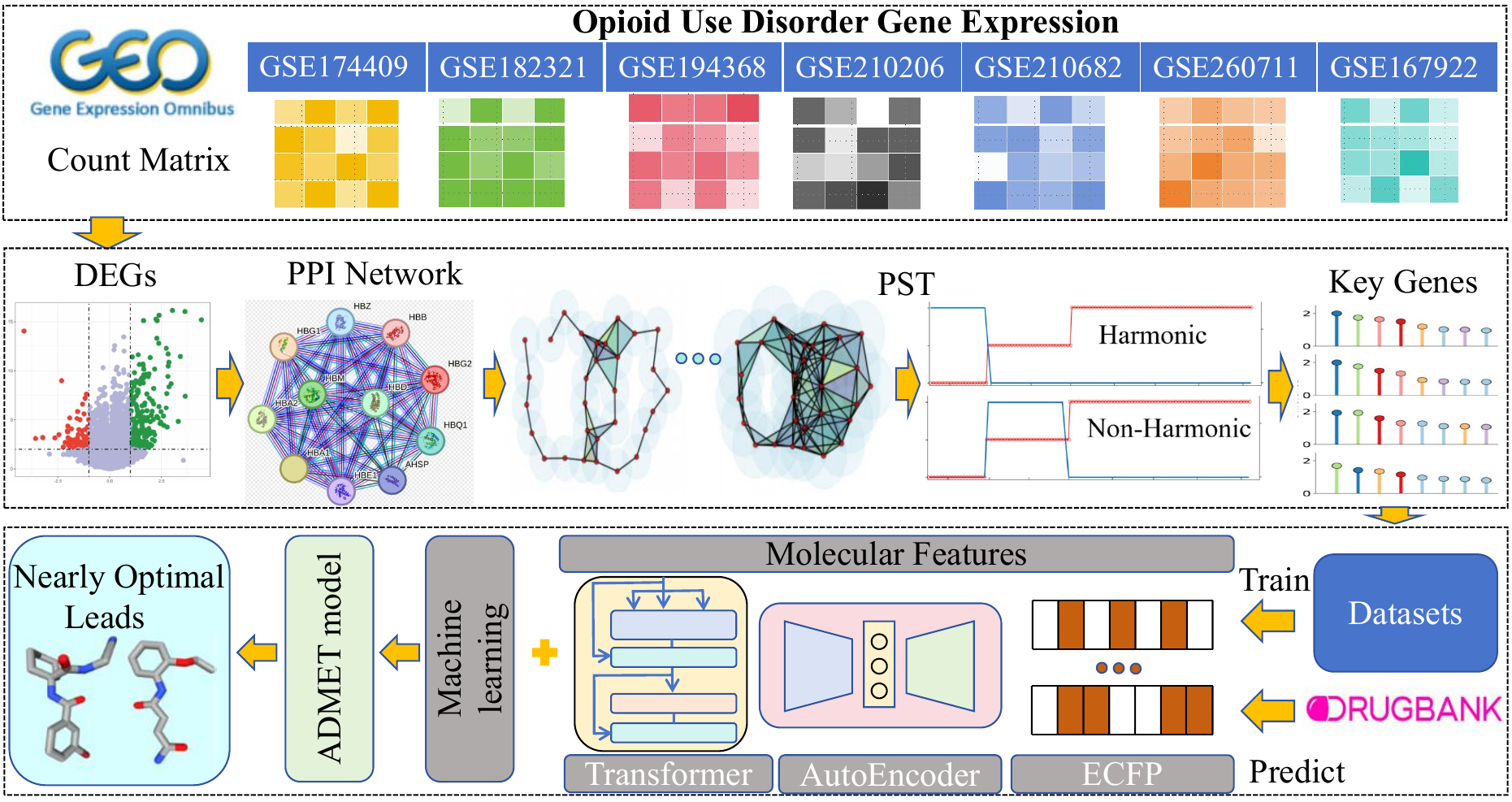} 
    \begin{minipage}{0.89\textwidth}
        \caption{ Illustration of the framework of the present study. Drawing on the GEO repository, we assembled count matrices of seven transcriptomic datasets linked to opioid addiction, and derived protein-protein interaction (PPI) networks from the resulting differentially expressed genes (DEGs) by differential expression profiling. And then, the topological differentiation based on persistent homology (PH) and persistent spectral theory (PST) was employed to extract key genes with strong mechanistic relevance to addiction. Using these genes as therapeutic targets, we retrieved corresponding inhibitor data from ChEMBL database to train  machine learning (ML) model. Three different ML algorithms in conjunction with three distinct molecular features was subsequently leveraged to systematically repurpose agents cataloged in DrugBank. All prioritized compounds underwent comprehensive ADMET (absorption, distribution, metabolism, excretion, and toxicity) characterization to confirm their drug-like properties. Finally, the nearly optimal lead compounds were screened.     
}
        \label{FIG1_flowchart}
    \end{minipage}
\end{figure*}

In this work, we present an innovation method to unearth potential drug repurposing candidates for opioid addiction treatment, aiming to bridge the gap between transcriptomic data analysis and drug discovery. The framework of this study is illustrated in Fig. \ref{FIG1_flowchart}. Specifically, we introduce a meta-analysis of seven transcriptomic datasets from the GEO database integrating with multiscale topological differentiation to identify key genes implicated in OUD. Following a rigorous validation through pathway analysis and data-availability scrutiny, we identify 1865 targets highly pertinent to opioid addiction for DrugBank repurposing. We developed ML models employing two natural language processing (NLP)-based embeddings generated via transformer and autoencoder models, as well as a traditional 2D fingerprint Extended Connectivity Fingerprint (ECFP), and got reliable predictive results in ten-fold cross-validation tests. Based on these ML models, we systematically evaluated the binding affinities of DrugBank compounds to those targets across distinct binding thresholds. This evaluation led to the identification of several drugs exhibiting satisfactory binding energies at specified binding affinity thresholds. Subsequently, molecular docking was performed on a select group of promising drugs to elucidate their interactions with receptors. Additionally, we have conducted a thorough ADMET (absorption, distribution, metabolism, excretion, and toxicity) analysis, providing a comprehensive evaluation of the pharmacokinetic and safety profiles of potential therapeutic compounds. The drugs identified, with their potent receptor inhibition affinities and favorable ADMET profiles, are prime candidates for subsequent biological experiment. This study not only identifies robust, network-validated targets but also prioritizes repurposable drugs with high binding affinities and favorable drug-likeness profiles. Ultimately, this approach seeks to accelerate the translation of genomic insights into actionable therapeutic strategies, offering a scalable model that can be applied to other complex diseases beyond opioid addiction.

\section{Results and Discussion} \label{sec:Results and Discussion}

\subsection{Opioid addiction meta-analysis} 

\subsubsection{Differential expression analysis} 
We systematically identified human transcriptomic datasets from the GEO database, focusing on studies investigating opioid addiction published within the past five years. The selected datasets (GSE174409, GSE182321, GSE194368, GSE210206, GSE210682, GSE260711, GSE167922) comprise literature-validated comparisons of gene expression profiles between opioid users and drug-naïve controls. While opioid analgesics remain clinically valuable due to their potent efficacy and administration convenience \cite{hwang2025evolution}, chronic use frequently leads to dependence and abuse, representing a significant public health challenge.

\begin{figure*}[htp]
    \centering
    \includegraphics[width=14cm]{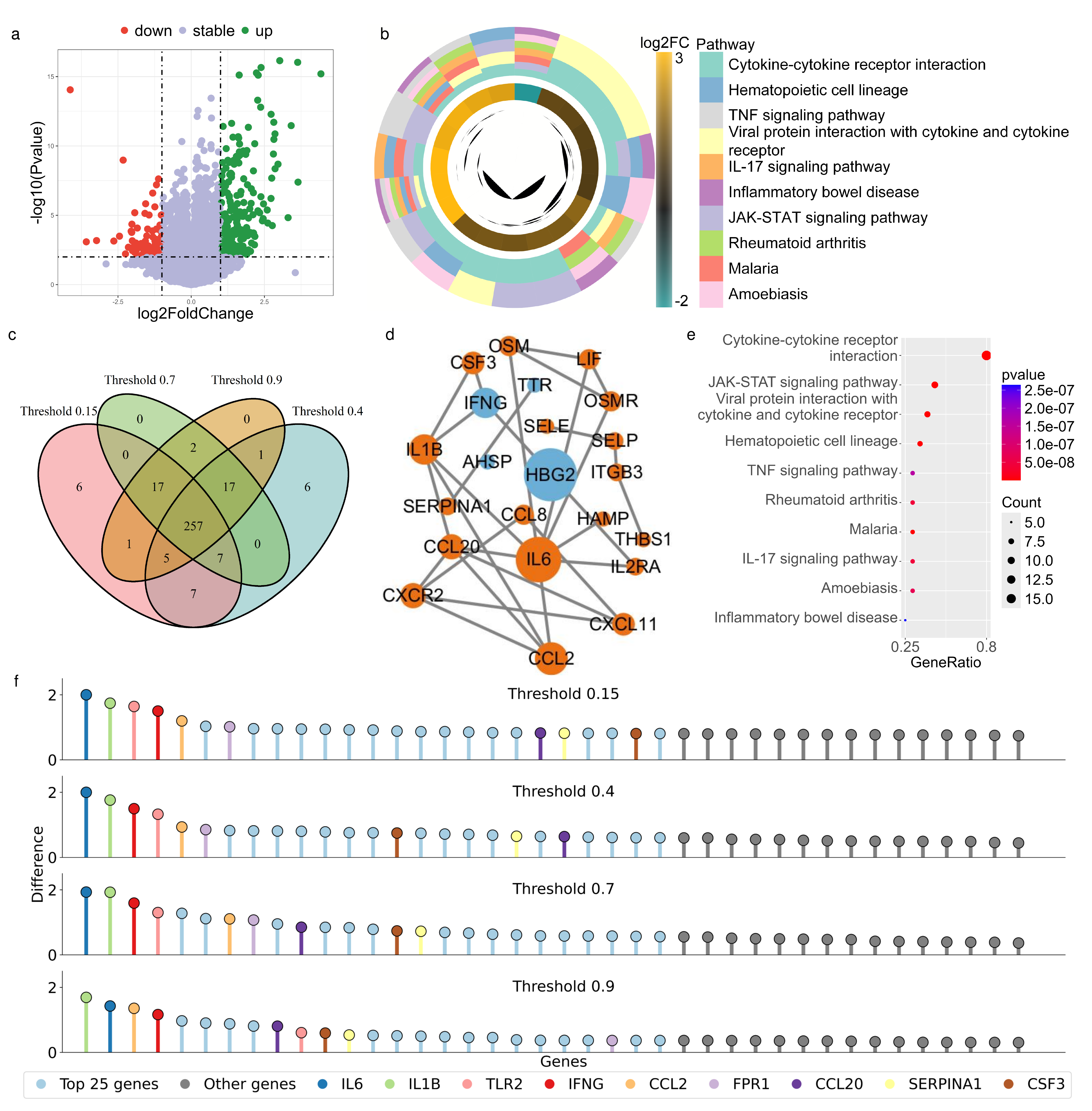} 
    \begin{minipage}{0.89\textwidth}
        \caption{The differentially expressed genes (DEGs) analysis for opioid addiction on GSE174409 dataset. \textbf{a}: The volcano plot of DEGs with red and green markers denoting down-regulated and up-regulated genes, respectively. This visualization effectively demonstrates both the statistical significance (\(-\log{10}(\text{Pvalue})\)) and biological relevance (\(\log{2}\text{FoldChange}\), or \(\log{2}\text{FC}\)) of gene expression alterations.
\textbf{b}:  Pathway enrichment analysis identifies the top 10 significantly enriched pathways associated with DEGs, suggesting their potential involvement in opioid addiction. The color gradient, corresponding to \(\log{2}\text{FC}\) values, highlights pathways with more pronounced expression changes, offering insights for investigating their biological roles.
\textbf{c}:  Venn diagram analysis demonstrates the overlap among the top 300 key genes identified at four distinct protein-protein interaction (PPI) network thresholds, pinpointing consistently important genes across varying stringency conditions.
\textbf{d}:  The PPI subnetwork depicts crucial molecular interactions among key genes, revealing potential functional relationships and central players in opioid addiction mechanisms.
\textbf{e}:  The bubble chart quantitatively represents pathway significance, where bubble size corresponds to gene count and color intensity reflects statistical significance (\(\text{Pvalue})\)), enabling visual comparison of pathway involvement.
\textbf{f}:  The PST-based (persistent spectral theory) network topological differentiation plot ranks gene importance, displaying the top 40 most significant genes to facilitate identification of key topological regulators in the biological network.        
}
        \label{FIG2_GSE174409}
    \end{minipage}
\end{figure*}

Differentially expressed gene (DEG) analysis was performed across all datasets using DESeq2 and Seurat method depending on the data format. The GSE174409 dataset exhibited 327 DEGs, with 233 up-regulated and 94 down-regulated as shown in Fig. \ref{FIG2_GSE174409}a where red color represents down regulated genes and green color represents up regulated genes. Analysis of GSE182321 revealed 314 DEGs (279 up-regulated, 35 down-regulated), while GSE194368 showed a similar pattern with 310 DEGs (277 up-regulated, 33 down-regulated). Notably, GSE210206 displayed the strongest up-regulation bias among all datasets, with 394 DEGs (373 up-regulated versus only 21 down-regulated). In contrast, GSE210682, while containing an identical number of total 394 DEGs as GSE210206, demonstrated an inverse expression pattern (131 up-regulated versus 263 down-regulated). The most substantial transcriptional changes were observed in GSE260711, which contained 442 DEGs (142 up-regulated and 300 down-regulated). Finally, GSE167922 analysis identified 314 DEGs with predominant up-regulation (214 up-regulated, 100 down-regulated). The details of DEGs analysis of these seven datasets can be found in Table \ref{DEG_analysis}. The DEG analysis offers us a preliminary insight into the possible molecular mechanisms behind opioid addiction, given that genes with differential expression can indicate pathways and processes modified in the disease condition.

\begin{table}[htbp]
\centering
\caption{Number and up/down-regulated distribution of DEGs of seven GEO datasets related to opioid addiction.}
\label{DEG_analysis}
{\scriptsize
\begin{tabular}{|>{\centering\arraybackslash}m{2.5cm}|>{\centering\arraybackslash}m{3cm}|>{\centering\arraybackslash}m{3cm}|>{\centering\arraybackslash}m{4cm}|}
\hline
Datasets\_name & Total\_DEGs\_number & Up-regulated\_number & Down-regulated\_number \\ \hline
GSE174409 & 327 & 233 & 94 \\ \hline
GSE182321 & 314 & 279 & 35 \\ \hline
GSE194368 & 310 & 277 & 33 \\ \hline
GSE210206 & 394 & 373 & 21 \\ \hline
GSE210682 & 394 & 131 & 263 \\ \hline
GSE260711 & 442 & 142 & 300 \\ \hline
GSE167922 & 314 & 214 & 100 \\ \hline
\end{tabular}
}
\end{table}

\begin{figure*}[htp]
    \centering
    \includegraphics[width=15cm]{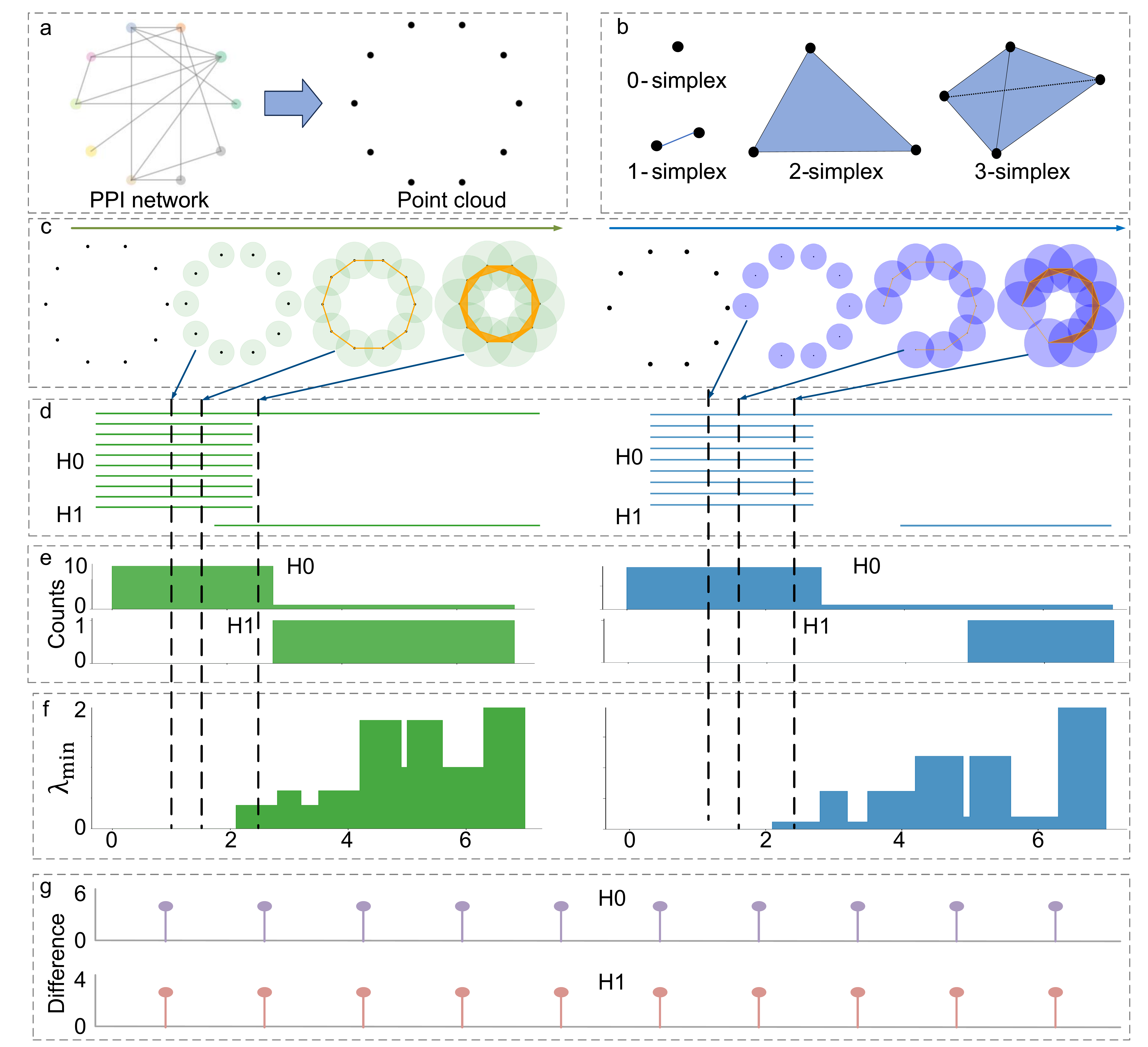} 
    \begin{minipage}{0.89\textwidth}
        \caption{ Schematic diagram of topological differentiation of network.
\textbf{a}: This panel illustrates the point cloud abstract representation of the protein-protein interaction (PPI) network, where the PPI network is simplified into a point cloud structure.  
\textbf{b}: Four basic simplx. They are the basic constituent units of simplicial complex.
\textbf{c}: The principle of the filtration or filtering process. The left side (green part) of this panel shows the simplicial complex evolution process of the original point cloud structure as the radius parameter gradually increases. The right side (purple part) indicates the newly formed network point cloud structure and its corresponding simplicial complex formed after removing one specific protein node, aiming to compare the impact of node deletion on the network topology structure.
\textbf{d}: Persistent barcode image. This panel uses persistent homology (PH) method to quantitatively describe the generation and disappearance of topological invariants in the filtration process of the network, thereby capturing the changes in the network's topological structure.
\textbf{e}: Changes in the count of topological invariants during the filtration process. This panel uses the persistent spectral theory (PST) method to track and display the spectra of persistent Laplacians, whose harmonic part corresponds to topological persistence.
\textbf{f}: The variation of the minimum value of the non-harmonic spectra. This panel highlights the evolution characteristics of homotopic shapes in the filtration process of data by analyzing the minimum values of non-harmonic spectra in PST.
\textbf{g}: The difference of topological invariants after node deletion in the network. The sum of changes during the filtration is given and the nodes on the far left mean the most significant changes. }
        \label{FIG3_TP}
    \end{minipage}
\end{figure*}

\subsubsection{Multiscale topological differentiation of PPI networks} 

To elucidate potential functional relationships among the identified DEGs, we employed the STRING database (version 11.5) to generate protein-protein interaction (PPI) networks across a range of interaction confidence thresholds (0.15, 0.4, 0.7, and 0.9). The STRING database synthesizes protein interaction data from experimental evidence, computational predictions, and established biological knowledge, providing standardized scoring to construct comprehensive interaction maps spanning physical and functional associations across multiple species \cite{szklarczyk2025string}. This multi-threshold analytical strategy facilitates the identification of both robust interactions and biologically relevant weaker associations. Fig. \ref{FIG2_GSE174409}d presents a PPI subnetwork constructed from DEGs of the GSE174409 dataset.

For network characterization, we applied the multiscale topological differentiation method developed by Du et al. \cite{du2024multiscale}, which integrates persistent homology (PH) and persistent spectral theory (PST) to quantify network topological and geometric properties, and the schematic diagram of its principle is shown in Fig. \ref{FIG3_TP}. The nodes in a PPI network can be simply seen as point cloud as shown in Fig. \ref{FIG3_TP}a, and four basic simplex examples are given in Fig. \ref{FIG3_TP}b.
By constructing simplicial complexes based on point cloud under different filtration or filtering radii, Fig. \ref{FIG3_TP}c presents the simplicial complexes constructed before and after the deletion of one node. Through the filtration process, Figs. \ref{FIG3_TP}d and \ref{FIG3_TP}e respectively reveal the persistent barcode and the counts of topological invariants corresponding to specific filtration radii, while Fig. \ref{FIG3_TP}f shows the change in the minimum of non-harmonic spectra during the filtration process. This technology can accurately extract the dynamically changing interaction patterns in PPI networks.

The core of this method lies in its quantitative filtration mechanism, which is based on the spatial distance between proteins derived from the confidence score of each interaction pairs. This precise and measurable analytical approach is significantly different from the traditional paradigm that mainly relies on centrality algorithms, which often fail to consider these quantitative interaction information. Our technical practice effectively demonstrates the potential application of topological data analysis (TDA), which can efficiently track the dynamic changes of topological invariants and capture the evolution process of homotopic structure. A significant feature of this method is its ability to dynamically parse network structures. Specifically, by systematically removing specific proteins and monitoring the resulting topological and geometric feature changes, we were able to evaluate the architecture stability of the network and measure the critical role of individual proteins. This type of analysis enables us to identify key proteins whose modifications or removals can significantly affect the overall structure of the network.

We employ PST to comprehensively analyze persistent Laplacian spectra across different networks. This analytical approach involves two key computational phases: initial extraction of topological persistence data from harmonic spectra, followed by enhanced characterization of homotopic shape evolution through non-harmonic spectral information. These extracted features undergo vectorization to establish a complete topological representation, enabling quantitative assessment of network structural changes. Specifically, Euclidean distance between vectorized features is computed before and after a targeted protein removal, serving as a metric to evaluate protein-specific network perturbations and determine nodal importance within the network. 

Based on this method, we identified nine key genes of the GSE1774409 dataset that had significant importance across the networks in Fig. \ref{FIG2_GSE174409}f: IL6, IL1B, TLR2, IFNG, CCL2, FPR1, CCL20, SERPINA1, and CSF3. All protein nodes within the networks were ranked by their nodal importance. The fact that they were consistently present across various network thresholds emphasizes their potential to play a central role in addiction-related biological pathways. Fig. \ref{FIG2_GSE174409}c showed in detail the intersection among the identified top 300 key genes under four different threshold values on the GSE174409 dataset.

As there are few shared key genes identified in all seven datasets, in order to obtain sufficient inhibitor datasets for these key genes, we extracted the top 300 genes from each dataset as key genes. The details of these key genes of seven datasets can be found in Table 1 of the Supporting Information. Initial aggregation across all seven datasets yielded 2100 candidate genes, which resulted in a final set of 1865 unique genes for subsequent analysis after removal of 235 redundant entries presented in multiple datasets. This refined gene set represents the core interactome components most likely to be involved in opioid addiction-related pathways. The other reason for choosing top 300 genes is that in the downstream binding affinities prediction, there needs enough inhibtor datasets to ensure the reliability of machine learning model. We tried to choose top 25, 100 or 200 genes from each dataset, however, the resulting final unique key genes have few inhibor dataset in ChEMBL database. So the top 300 genes is set to identify the key genes.

\subsubsection{Pathway enrichment analysis on the GSE174409 dataset}

To further investigate the biological mechanisms of these key genes in opioid addiction, we conducted a pathway enrichment analysis, which showed that multiple signaling pathways exhibited significant enrichment in DEGs of the GSE174409 dataset. Specifically, Figs. \ref{FIG2_GSE174409}b and \ref{FIG2_GSE174409}e describe the top 10 enriched signaling pathways in the dataset. Among them, several pathways are particularly prominent, such as the cytokine-cytokine receptor interaction pathway, JAK-STAT (Janus kinase-signal transducers and activators of transcription) signaling pathway, and the viral protein interaction with cytokine and cytokine receptor pathway, which contain 16, 9, and 8 DEGs, respectively.

Opioid addiction occurs through multiple signaling pathways. Firstly, morphine exhibits a dual effect in the cytokine-cytokine receptor interaction pathway: on the one hand, it can trigger neuroinflammatory responses in the central nervous system, which have been shown to be a key driving factor in enhancing drug dependence\cite{goldstein202242}. On the other hand, morphine can also suppress the peripheral immune system as a whole, manifested by a decrease in the number of immune cells and downregulation of pro-inflammatory cytokines and chemokines expression \cite{qiu2023estradiol,malik2024morphine}. Secondly, at the level of Hematopoietic cell lineage, opioid addiction has been shown to have direct or indirect toxicity or functional interference on hematopoietic stem/progenitor cells \cite{quraishi2022effect}. In the TNF signaling pathway, activation of the kappa opioid receptor (KOR) can weaken TNF-$\alpha$-mediated inflammatory response and cartilage degradation by inhibiting the STAT3 cascade \cite{liu2024activation}. Additionally, studies on the viral protein interaction with cytokine and cytokine receptor suggest that long-term exposure to opioids weakens innate and adaptive immunity, significantly increasing host susceptibility to multiple pathogens. One of the mechanisms involved is the systemic inhibition of macrophage function \cite{wen2022opioids,roy2011opioid}. KOR agonists have been shown to improve the cognitive status of rats with cognitive impairment after nerve injury by inhibiting the JAK2/STAT3 signaling pathway \cite{li2019kappa}. Furthermore, a whole transcriptome sequencing study published in 2022 also found significant enrichment of JAK-STAT pathway related genes in peripheral blood samples of patients with opioid use disorders, and abnormal expression of cytokines closely related to immune regulation, such as IL-6 and interferon \cite{dai2022whole}.

\subsection{Repurposing of DrugBank for opioid addiction targets} 

\subsubsection{Binding affinity predictors for opoiod addiction targets}

To investigate potential therapeutic agents for opioid addiction, we performed drug repurposing analysis on compounds from the DrugBank database using ML approaches. Since ML model performance depends critically on data quality and quantity, we retrieved corresponding inhibitor data from the ChEMBL database based on the previously identified 1,865 key genes. To ensure data quality and meet the required sample size for model training, we have set the following screening criteria: the number of compounds in each dataset must be greater than or equal to 250 to ensure sufficient training samples. In addition, it must also meet the requirements of individual proteins and human genes.

This selection process yielded 76 inhibitor datasets, however, four datasets (CHEMMBL3989381, CHEMMBL4506, CHEMMBL4105860, and CHEMMBL3544) were subsequently excluded due to insufficient compound counts that could compromise model training efficacy. Hence, the final analysis incorporated 72 datasets comprising 46,977 compounds. The information of ID name and sample size of these 72 datasets can be found in Table 2 of the Supporting Information.

For effective molecular structure representation used in ML models, we employed multiple molecular representation approaches. Specifically, we combined deep learning techniques with conventional molecular fingerprinting methods, utilizing 
sequence-to-sequence autoencoders fingerprints (AE-TP), bidirectional transformer fingerprints (BET-TP), along with a traditional  fingerprints (ECFP) to capture detailed molecular structural features. These derived features were subsequently used for ML model training. We implemented three distinct ML algorithms: support vector machine (SVM), random forest (RF), and gradient boosting decision tree (GBDT) to predict drug-target binding affinities, with detailed parameter specifications provided in Table 3 of the Supporting Information.

During model training, in order to obtain the reliable prediction results, we combined three ML algorithms (SVM, RF, and GBDT) and three molecular fingerprints (AE-TP, BET-TP, and ECTP), totaling 21 distinct model configurations. Model performance was assessed using three evaluation metrics: Pearson correlation coefficient (P), coefficient of determination ($R^2$), and root mean square error (RMSE). Comparative analysis demonstrated that the model integrating SVM regressor with three fingerprints fused outperformed other model configurations on most 72 inhabitor datasets. The complete predictive results can be found in Tables  4, 5, and 6 of the Supporting Information.

\subsubsection{Potential inhibitors of opioid addiction targets in DrugBank} 

To identify potential inhibitors targeting opioid addiction proteins, we employed ML models to predict binding affinities of small molecules from the DrugBank database. DrugBank categorizes small molecules according to their clinical development status, and our study specifically focused on compounds classified as either "Approved" or "Investigational". This selection was based on two  points: approved drugs possess established safety profiles and clinical validation, facilitating rapid repurposing opportunities, while investigational compounds, though not yet marketed, typically have supporting clinical data that may reveal novel therapeutic applications. Together, these two categories encompassed 6,001 small molecules for analysis. 

In order to ensure the reliability of the prediction results during the implementation process, we adopted the 10-fold cross-validation method. Furthermore, to prioritize high-affinity candidate molecules, we implemented a binding affinity (BA) threshold of -9.54 kcal/mol, retaining only those compounds with predicted values exceeding this cutoff. This stringent criterion is well-established and widely adopted in related research domains, ensuring the biological relevance of our screening results \cite{flower2002drug}.

\paragraph{Approved drugs with predicted efficacy on GPR84}

Among the 300 key genes identified for the GSE174409 dataset, only GPR84 and F2RL3, whose corresponding inhibitor dataset are ChEMBL3714079 and ChEMBL4691, respectively, satisfied our predefined inhibitor screening criteria. We subsequently evaluated approved drugs exhibiting strong BA (predicted binding free energy $\le$ -9.54 kcal/mol) against these targets. Table \ref{tab:GPR84_approved} presents the top 15 representative drugs with strongest binding affinities for target GPR84. The discussion about the first three drugs, estradiol cypionate, bosentan, and givinostat is given in the following.

\begin{table}[htbp]
\centering
\caption{Summary of the FDA-approved drugs that are potential potent inhibitors of GPR84 with binding affinity (BA) smaller than -9.54 kcal/mol.}
\label{tab:GPR84_approved}
{\scriptsize
\begin{tabular}{|>{\centering\arraybackslash}m{2.5cm}|>{\centering\arraybackslash}m{4.5cm}|>{\centering\arraybackslash}m{4cm}|}
\hline
DrugBank ID & Generic Name & Predicted BA (kcal/mol) \\ \hline
DB13954 & Estradiol cypionate & -12.34 \\ \hline
DB00559 & Bosentan & -11.72 \\ \hline
DB12645 & Givinostat & -11.72 \\ \hline
DB12301 & Doravirine & -11.71 \\ \hline
DB13211 & Guanoxan & -11.71 \\ \hline
DB00590 & Doxazosin & -11.69 \\ \hline
DB01179 & Podofilox & -11.69 \\ \hline
DB11637 & Delamanid & -11.69 \\ \hline
DB09299 & Tenofovir alafenamide & -11.69 \\ \hline
DB06811 & Polidocanol & -11.68 \\ \hline
DB06119 & Cenobamate & -11.68 \\ \hline
DB00509 & Dextrothyroxine & -11.67 \\ \hline
DB01599 & Probucol & -11.67 \\ \hline
DB01117 & Atovaquone & -11.66 \\ \hline
DB00654 & Latanoprost & -11.66 \\ \hline
\end{tabular}
}
\end{table}

Estradiol cypionate (ECP) is a synthetic estrogen derivative produced via esterification of estradiol's 17$\beta$-hydroxyl group with cyclopentylpropionic acid. This long-acting estrogen formulation was first approved for clinical use in 1952 to manage symptoms of menopausal ovarian failure \cite{goldstein202242}. Beyond traditional hormone replacement applications, emerging research indicates ECP's potential anticancer properties. In vitro studies demonstrate ECP significantly inhibits lung cancer cell proliferation \cite{du2024rolapitant}. Especially, it has been found that ECP can inhibit gastric cancer growth and induce cell apoptosis by regulating the PI3K/Akt/mTOR signaling pathway \cite{qiu2023estradiol}. It is worth noting that this pathway is also closely related to opioid addiction \cite{singh2025pharmacological}. This association suggests that the regulatory effect of ECP on the PI3K/Akt/mTOR pathway may provide a potential biological basis for exploring its use as a treatment strategy for opioid addiction.

Givinostat is a small molecule histone deacetylase (HDAC) inhibitor that reduces inflammation and muscle loss by targeting pathogenic processes \cite{sandona2023histone}. It was first approved in the United States in 2024 for the treatment of muscular dystrophy and polycythemia vera \cite{lamb2024givinostat}.  The existing research suggests that HDAC inhibitors may help reduce dependence on opioid drugs \cite{pryce2022novel}. This potential is closely related to its mechanism of action: $\delta$-opioid receptor agonists have been shown to be associated with HDAC class I and IIb activity \cite{zaidi2020histone}. Therefore, givinostat, as an HDAC inhibitor, has not only established its clinical application value, but also provided a promising research direction for exploring new strategies to address opioid addiction, and has important potential clinical significance.

Bosentan is an oral dual endothelin receptor antagonist (endothelin-A and endothelin-B) \cite{ono2002endothelin}. Its clinical application initially focused on the cardiovascular system, used to treat pulmonary arterial hypertension, by antagonizing the action of endothelin (ET-1), blocking the vasoconstriction and hypertensive effects induced by this molecule \cite{enevoldsen2020endothelin}. Additionally, recent studies have found that bosentan can enhance the inhibitory effect of morphine on thermal and tactile hypersensitivity in tumor induced pain models, and has an opioid saving effect \cite{kopruszinski2018blockade}. Meanwhile, as an ETA/ET-B dual receptor antagonist, bosentan can enhance the analgesic effect of opioids and eliminate analgesic tolerance, providing potential new ideas for developing novel interventions for opioid addiction \cite{bhalla2016neurobiology}.

\begin{figure*}[htp]
    \centering
    \includegraphics[width=13cm]{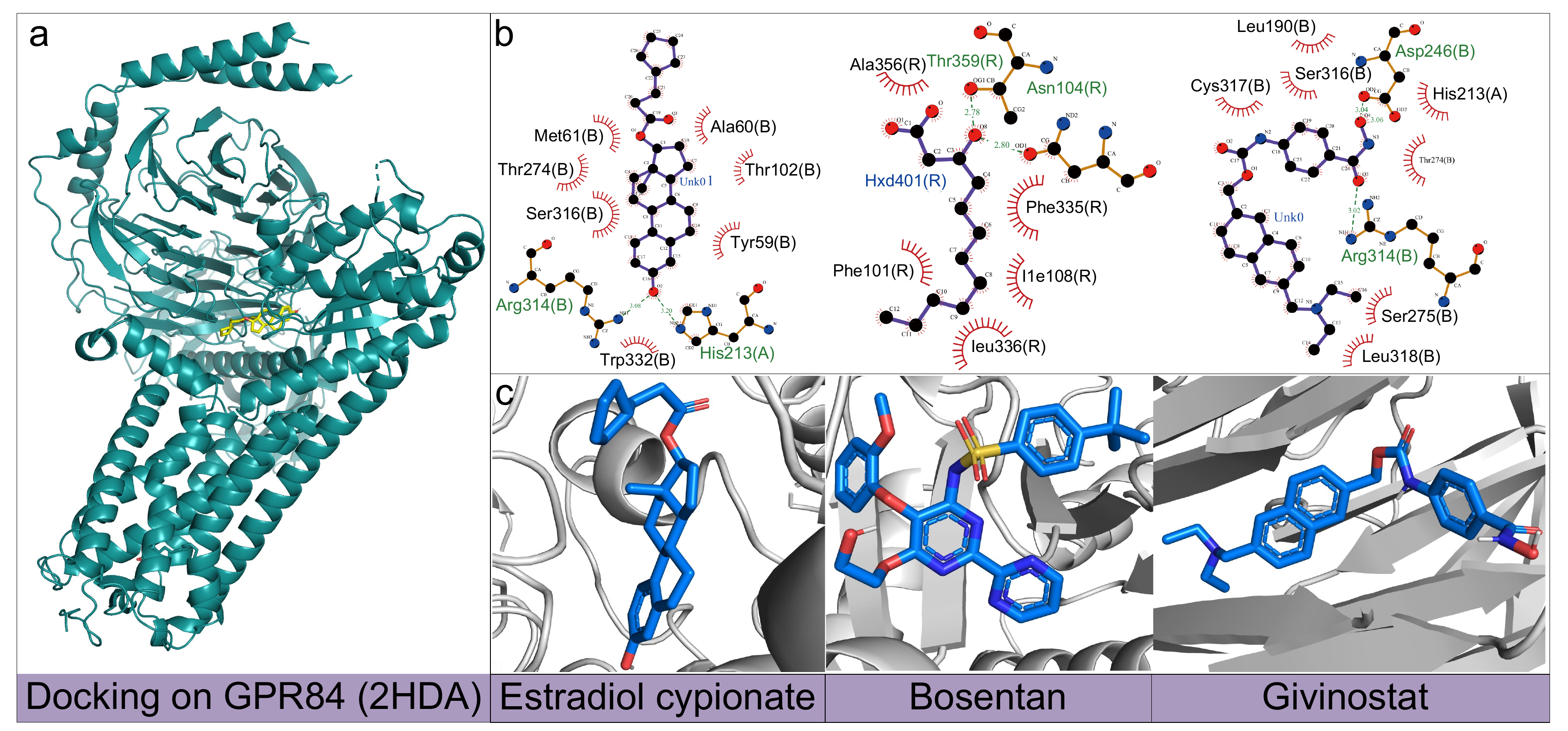} 
    \begin{minipage}{0.89\textwidth}
        \caption{Docking conformations and interactions of estradiol cypionate, bosentan, and givinostat with GPR84 (2HDA). 
        \textbf{a}: The three-dimensional (3D) molecular docking conformation of estradiol cypionate complexed with the GPR84 (2HDA). 
        \textbf{b}: The two-dimensional interaction diagram between the three drugs and GPR84 (2HDA). 
        \textbf{c}: The 3D molecular configurations of the three drugs.}
        \label{FIG4_GPR84}
    \end{minipage}
\end{figure*}

To investigate the interaction patterns between these three drugs and the GPR84 receptor, we conducted molecular docking studies using AutoDock Vina software. Specifically, we docked estradiol cypionate, bosentan, and givinostat with GPR84 protein (PDB ID: 2HDA), respectively. Nine candidate binding conformations or poses for each drug were generated and their corresponding binding energies (in kcal/mol) based on their scoring functions were calculated. The details of docking between three drugs and GPR84 can be found in Tables 7, 8, and 9 of the Supporting Information. In the analysis and visualization of the docking results, we chose the conformation with the lowest binding energy (i.e. the highest affinity) for display. Fig. \ref{FIG4_GPR84} presents the molecular docking results of three drugs with GPR84 protein. Specifically, Fig. \ref{FIG4_GPR84}a shows the docking conformation of GPR84 with estradiol cypionate. Fig. \ref{FIG4_GPR84}b displays the two-dimensional interaction diagrams of these three drugs, and Fig. \ref{FIG4_GPR84}c presents the three-dimensional (3D) structures of these drugs. The docking results indicated that all three drugs formed two hydrogen bonds with GPR84 protein. It is worth noting that both estradiol cypionate and givinostat form hydrogen bonds with the Arg314 residue, which may suggest that Arg314 plays a key role in mediating the interaction between these drugs and GPR84.

\paragraph{Investigational drugs with predicted efficacy on GPR84}

\begin{figure*}[htp]
    \centering
    \includegraphics[width=15cm]{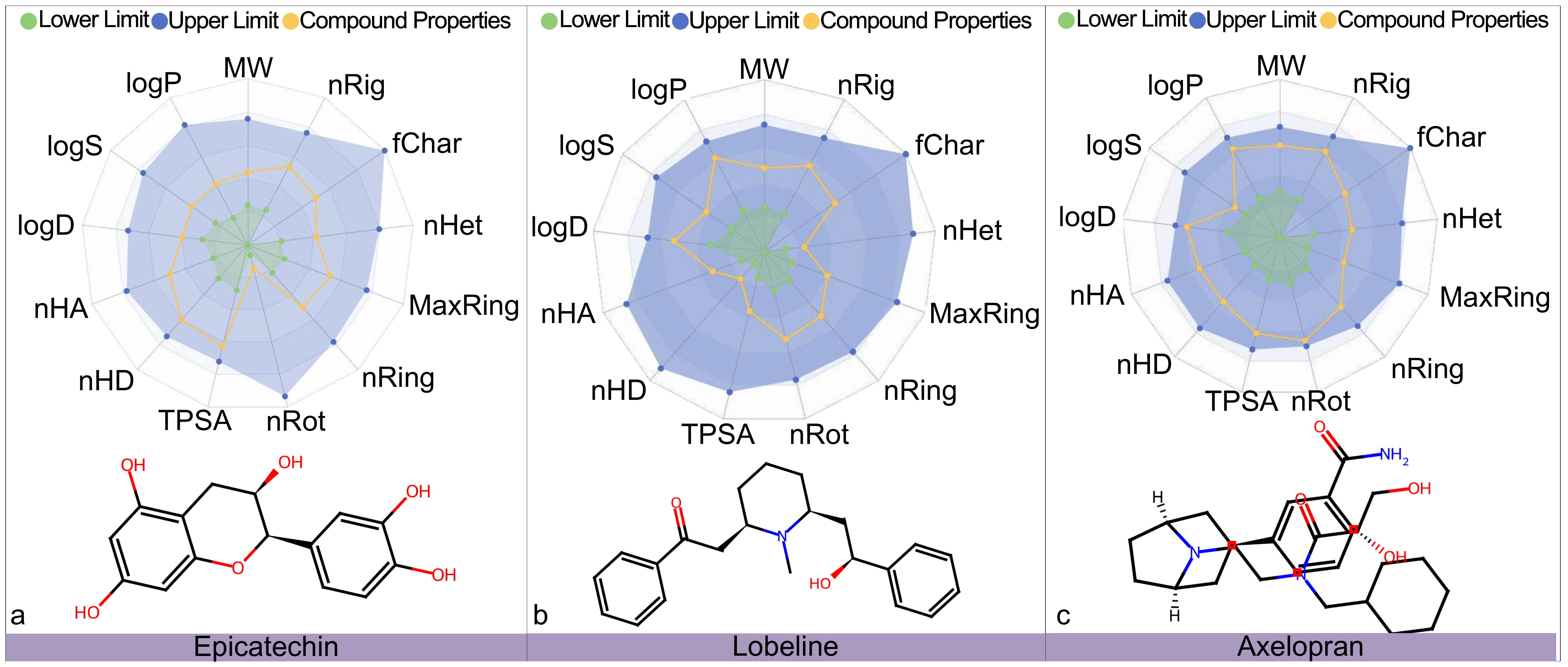} 
    \begin{minipage}{0.89\textwidth}
        \caption{ ADMET (absorption, distribution, metabolism, excretion, and toxicity) property evaluation for epicatechin,  lobeline, and axelopran. The blue zone defines the upper boundary of optimal ranges for 13 specified ADMET properties, while the green zone indicates the lower boundary. The yellow curve depicts drug-specific property values. Predictive data were retrieved from the ADMETlab 3.0 platform (https://admetlab3.scbdd.com/). The evaluated ADMET properties include MolecularWeight (MW), the logarithm of the octanol/water partition coefficient (logP), the logarithm of the aqueous solubility (logS), the logP at physiological pH 7.4 (logD), Number of hydrogen bond acceptors (nHA), Number of hydrogen bond donors (nHD), Topological polar surface area (TPSA), Number of rotatable bonds (nRot), Number of rings (nRing), Number of atoms in the biggest ring (MaxRing), Number of heteroatoms (nHet), Formal charge (fChar), and Number of rigid bonds (nRig).}
        \label{FIG5_GPR84_ADMET}
    \end{minipage}
\end{figure*}

Table \ref{tab:GPR84_investigational} lists the top 15 investigational drugs from DrugBank ranked by predicted BA values. To comprehensively analyze their ADMET properties, we performed computational profiling using the ADMETlab 3.0 platform (https://admetlab3.scbdd.com/). The ADMET property is a key factor determining whether a candidate drug can successfully pass preclinical screening and enter the subsequent development stage. Fig. \ref{FIG5_GPR84_ADMET} illustrates the predicted ADMET profiles for top 3 drugs in Table \ref{tab:GPR84_investigational}, epicatechin, lobeline, and axelopran, respectively, which indicates that all three drugs exhibit ADMET properties within the optimal range.

\begin{table}[htbp]
\centering
\caption{Summary of investigational drugs that have the potential to inhibit GPR84.}
\label{tab:GPR84_investigational}
{\scriptsize
\begin{tabular}{|>{\centering\arraybackslash}m{2.5cm}|>{\centering\arraybackslash}m{4.5cm}|>{\centering\arraybackslash}m{4cm}|}
\hline
DrugBank ID & Generic Name & Predicted BA (kcal/mol) \\ \hline
DB12039 & Epicatechin & -11.20 \\ \hline
DB05137 & Lobeline & -11.20 \\ \hline
DB12013 & Axelopran & -11.12 \\ \hline
DB04903 & Pagoclone & -11.12 \\ \hline
DB12556 & MK-5108 & -11.12 \\ \hline
DB18252 & Pralmorelin & -11.11 \\ \hline
DB11745 & Otenabant & -11.11 \\ \hline
DB16037 & BI 44370 TA & -11.11 \\ \hline
DB16032 & GW810781 & -11.11 \\ \hline
DB17622 & KL1333 & -11.11 \\ \hline
DB16347 & Velsecorat & -11.11 \\ \hline
DB12724 & AZD-7295 & -11.11 \\ \hline
DB16068 & BTRX-335140 & -11.11 \\ \hline
DB12381 & Merestinib & -11.11 \\ \hline
DB18479 & Rodatristat ethyl & -11.11 \\ \hline
\end{tabular}
}
\end{table}

Epicatechin is a small molecule compound. Research has shown that epicatechin has the ability to bind to opioid receptors and activate downstream signaling mediated by opioid receptors, thereby exerting a regulatory effect on myocardial ischemia-reperfusion injury in vivo \cite{panneerselvam2010dark}. Additionally, the vascular response and cardioprotective effects of epicatechin are mediated through multiple mechanisms, including activation of opioid receptors, nitric oxide, potassium channels, and calcium channels \cite{macrae2019epicatechin}. It is worth noting that epicatechin has been studied in the early treatment trials of diabetes and cancer \cite{quinonez2013analysis,shay2015molecular}. Given its interaction with opioid receptors, epicatechin may have the potential to address opioid addiction, and this direction deserves further in-depth research.

Lobeline itself, as an alkaloid \cite{xu2022cis}, has a similar effect on nicotinic acetylcholine receptors as nicotine, although its potency is relatively weak. This characteristic makes lobeline potentially valuable in various therapeutic fields, including but not limited to the treatment of colon cancer \cite{zhao2025targeting} and neurological protection \cite{remya2023lobeline}. Additionally, recent study has shown that lobeline has the function of a $\mu$-opioid receptor antagonist, which makes it a promising effective treatment for opioid addiction \cite{miller2007lobeline}. Other studies related to lobeline have also revealed its potential application value in the field of neurobiology. For example, lobeline has been shown to bind to $\mu$-opioid receptors, block the effects of opioid receptor agonists, and significantly reduce heroin self-administration behavior in rats \cite{miller2011lobeline}. This mechanism of action provides a theoretical basis for the application of lobeline in the treatment of opioid addiction.

Axelopran is an experimental drug that has been used in the treatment of ovarian induced constriction \cite{longare2023state}. As a peripheral acting $\mu$-opioid receptor antagonist \cite{holder2016novel}, it theoretically has the potential to address opioid addiction and provide new solutions for the treatment of opioid addiction.

\paragraph{Approved drugs with predicted efficacy on F2RL3}

Table \ref{tab:F2RL3_approved} enumerates the top 15 FDA-approved drugs exhibiting potential F2RL3 inhibitory activity. 

\begin{figure*}[htp]
    \centering
    \includegraphics[width=13cm]{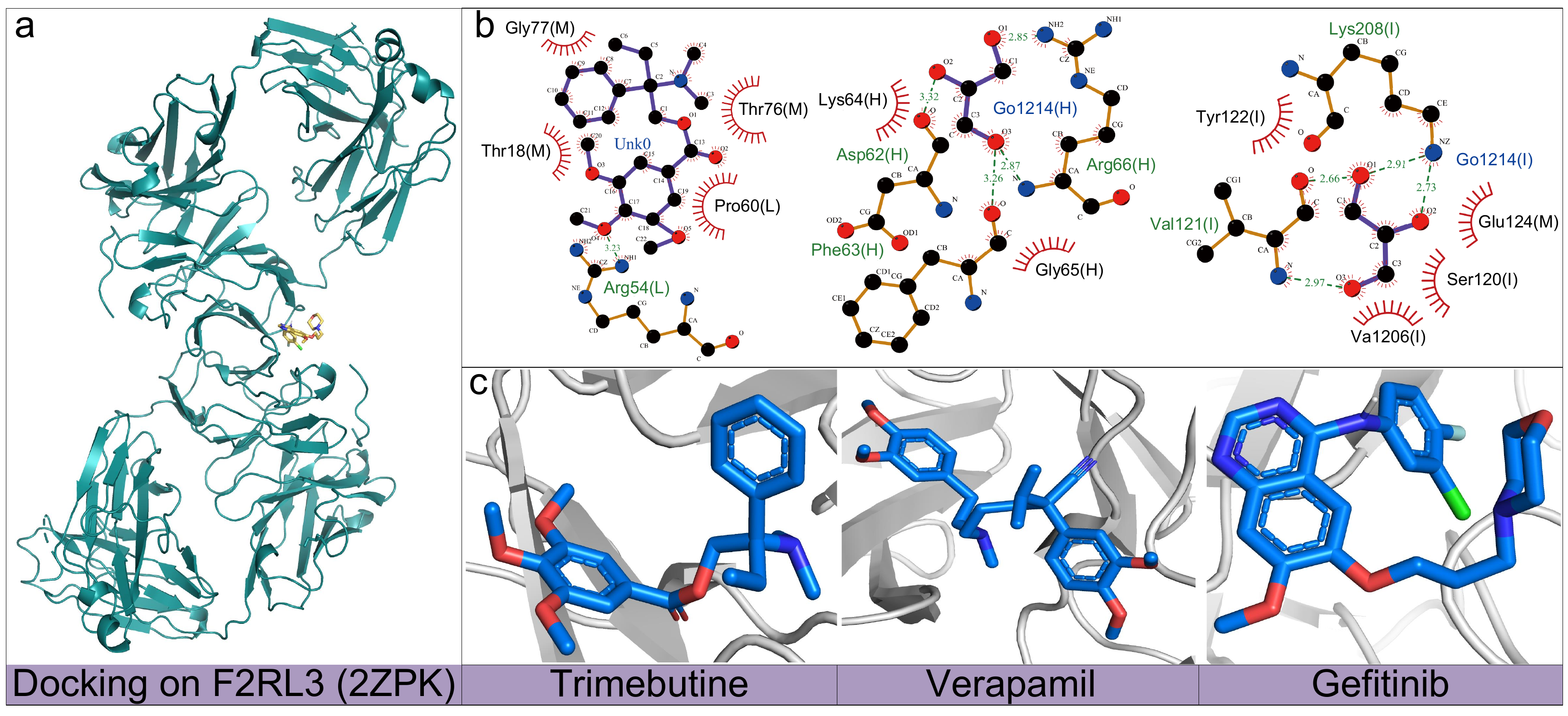} 
    \begin{minipage}{0.89\textwidth}
        \caption{Molecular docking structure and interaction of trimebutine, verapamil, gefitinib with F2RL3 protein (PDB ID: 2ZPK). \textbf{a}: This diagram showcases the spatial orientation of gefitinib when it binds to F2RL3 protein (2ZPK). \textbf{b}: The illustration depicts the planar representation of interactions between three drugs and F2RL3. \textbf{c}: The three-dimensional structural arrangements of three drugs.}
        \label{FIG6_F2Rl3}
    \end{minipage}
\end{figure*}

\begin{table}[htbp]
\centering
\caption{Summary of the FDA-approved drugs that are potential potent inhibitors of F2RL3 with binding affinity (BA) smaller than -9.54 kcal/mol.}
\label{tab:F2RL3_approved}
{\scriptsize
\begin{tabular}{|>{\centering\arraybackslash}m{2.5cm}|>{\centering\arraybackslash}m{4.5cm}|>{\centering\arraybackslash}m{4cm}|}
\hline
DrugBank ID & Generic Name & Predicted BA (kcal/mol) \\ \hline
DB09089 & Trimebutine & -11.19 \\ \hline
DB00661 & Verapamil & -11.05 \\ \hline
DB00317 & Gefitinib & -11.00 \\ \hline
DB11963 & Dacomitinib & -10.99 \\ \hline
DB06616 & Bosutinib & -10.99 \\ \hline
DB13277 & Benziodarone & -10.98 \\ \hline
DB00357 & Aminoglutethimide & -10.98 \\ \hline
DB15097 & Gefapixant & -10.96 \\ \hline
DB01113 & Papaverine & -10.95 \\ \hline
DB11611 & Lifitegrast & -10.95 \\ \hline
DB01016 & Glyburide & -10.94 \\ \hline
DB01581 & Sulfamerazine & -10.94 \\ \hline
DB09080 & Olodaterol & -10.93 \\ \hline
DB01118 & Amiodarone & -10.92 \\ \hline
DB11768 & Zytron & -10.92 \\ \hline
\end{tabular}
}
\end{table}

Trimebutine is not only a commonly used antispasmodic agent for symptomatic treatment of irritable bowel syndrome \cite{kountouras2002efficacy}, but also an effective method for treating functional gastrointestinal disease \cite{andreev2021efficacy}. However, recent studies have revealed its broader pharmacological effects. It has been found that trimethoprim differs from classical antispasmodics in that it has a weak but significant excitatory effect on intestinal opioid receptors ($\mu$ and $\kappa$) \cite{cenac2016novel}. As an opioid ligand, trimebutine can interact with $\mu$ , $\sigma$, and $\kappa$ receptor subtypes with similar affinity \cite{collins1991opiate}. Specifically, it exhibits a relatively high affinity for the $\mu$ receptor subtype \cite{roman1987interactions}.

Verapamil, as a phenylalkylamine calcium channel blocker \cite{weizman1999pharmacological}, was the first calcium channel antagonist to be clinically used in the early 1960s. Its main indications include hypertension \cite{mctavish1989verapamil} and diabetes \cite{borowiec2022txnip}. Recent studies suggest that verapamil may have potential therapeutic effects on acute opioid withdrawal syndrome, and further in-depth research is warranted \cite{salat2010comparative,ansari2003calcium}.

Gefitinib is an orally active selective epidermal growth factor receptor \cite{rawluk2018gefitinib}. This drug is mainly used to treat non-small cell lung cancer \cite{culy2002gefitinib}. Research has found that EGFR antagonist gefitinib effectively eliminates morphine tolerance and may have important clinical value in the treatment of neuropathic pain with opioid tolerance and poor response to opioid therapy \cite{puig2020egfr}.

To investigate the interaction mechanism between three drugs and F2RL3 protein, we used molecular docking simulation technology. As shown in Fig. \ref{FIG6_F2Rl3}, the docking results revealed the binding patterns of each drug with the target protein. Specifically, Fig. \ref{FIG6_F2Rl3}a illustrates the binding conformation of drug molecules. Fig. \ref{FIG6_F2Rl3}b provides a detailed description of the corresponding intermolecular interactions, and the 3D structures of the three compounds are shown in Fig. \ref{FIG6_F2Rl3}c. The results show that all three drugs can form stable hydrogen bonds with key amino acid residues of F2RL3 protein through the oxygen atoms in their molecular structures as hydrogen bond acceptors. Among them, the oxygen atom of trimebutine forms hydrogen bonds with the Arg54 residue. The oxygen atom of verapamil forms hydrogen bonds with Asp62, Phe63, and Arg66 residues simultaneously. Gefitinib forms hydrogen bonds by interacting with Val121 and Lys208 residues through its oxygen atom. The details of docking between these drugs and F2RL3 are given in Tables 10, 11, and 12 of the Supporting Information.

\paragraph{Investigational drugs with predicted efficacy on F2RL3}

Table \ref{tab:F2RL3_investigational} lists the top 15 drugs with a status of "investigational" and sorts them according to the predicted BA values.

\begin{figure*}[htp]
    \centering
    \includegraphics[width=15cm]{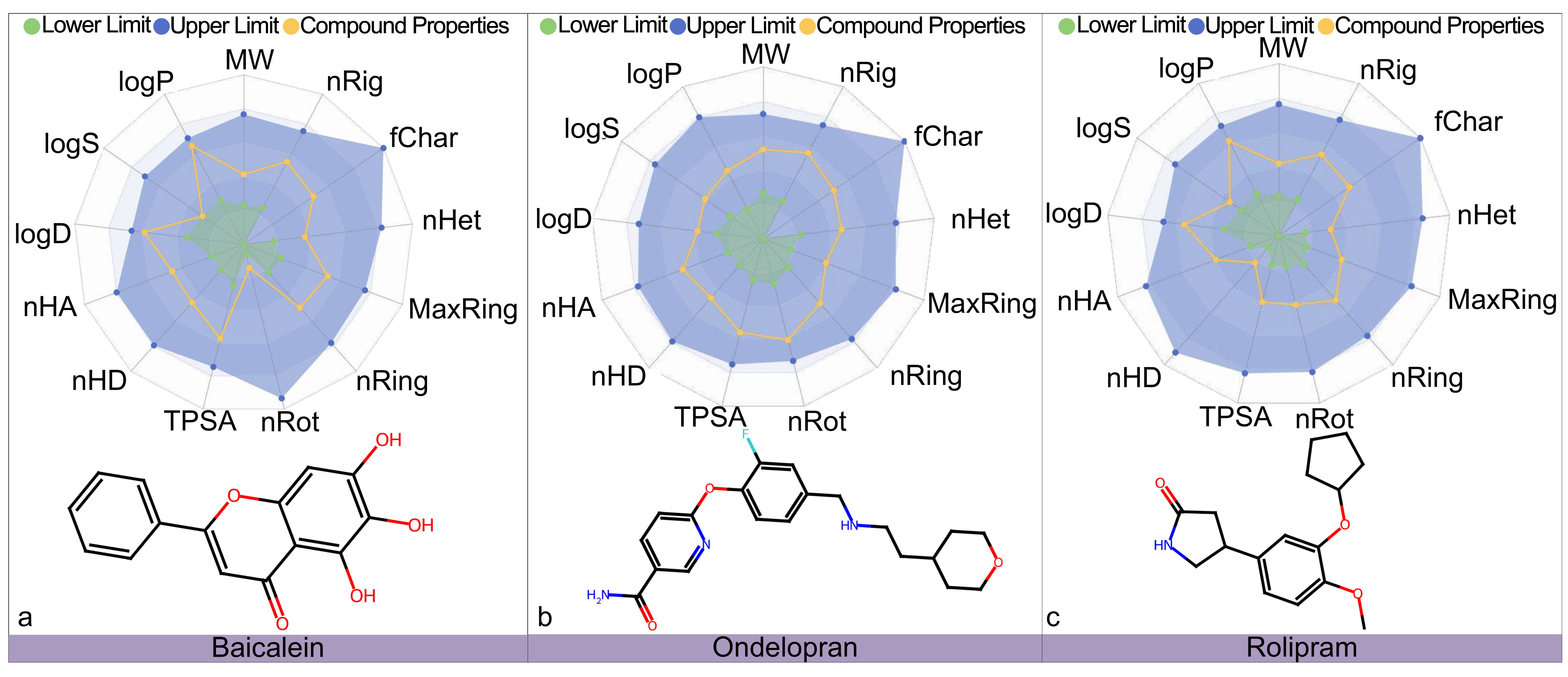} 
    \begin{minipage}{0.89\textwidth}
        \caption{This figure showcases the ADMET (absorption, distribution, metabolism, excretion, and toxicity) profiles for baicaleim, ondelopran, and rolipram. The yellow curves in each panel indicate the values for 13 specific ADMET
properties of these compounds. The blue and green zones in the graph are designated to highlight the upper and lower limits of the optimal ranges for each of these ADMET properties, respectively.}
        \label{FIG7_F2RL3_ADMET}
    \end{minipage}
\end{figure*}

\begin{table}[htbp]
\centering
\caption{Summary of investigational drugs that have the potential to inhibit F2RL3.}
\label{tab:F2RL3_investigational}
{\scriptsize
\begin{tabular}{|>{\centering\arraybackslash}m{2.5cm}|>{\centering\arraybackslash}m{4.5cm}|>{\centering\arraybackslash}m{4cm}|}
\hline
DrugBank ID & Generic Name & Predicted BA (kcal/mol) \\ \hline
DB01954 & Rolipram & -11.39 \\ \hline
DB16101 & Baicalein & -11.04 \\ \hline
DB12585 & Ondelopran & -10.98 \\ \hline
DB17102 & Ioflubenzamide I-131 & -10.98 \\ \hline
DB17050 & RS-39604 & -10.98 \\ \hline
DB18061 & AKV-9 & -10.97 \\ \hline
DB17199 & AAG-1 & -10.97 \\ \hline
DB00269 & Chlorotrianisene & -10.96 \\ \hline
DB05095 & Cimicoxib & -10.96 \\ \hline
DB17939 & Pterostilbene & -10.95 \\ \hline
DB02292 & Irosustat & -10.95 \\ \hline
DB03849 & Cilomilast & -10.95 \\ \hline
DB05546 & SB-743921 & -10.94 \\ \hline
DB15292 & Foliglurax & -10.94 \\ \hline
DB16042 & SB-773812 & -10.94 \\ \hline
\end{tabular}
}
\end{table}

Rolipram is a phosphodiesterase inhibitor with antidepressant activity. It may improve AD (Alzheimer's disease) related cognitive impairment and depressive like behavior by reducing amyloid beta, tau phosphorylation, neuroinflammation, and cell apoptosis. The mechanism may involve the cAMP/PKA/26S and cAMP/PACE/ERK pathways, suggesting that it may be a potential target for treating AD with depression \cite{cong2023rolipram}. Additionally, rolipram can effectively inhibit the downstream signal of Hedgehog (Hh) pathway. Given that the abnormality of this pathway is related to the progress and metastasis of breast cancer and other cancers \cite{bagchi2025pde4}, it is expected to be used to control the development of breast cancer. Furthermore, rolipram may help alleviate the development of morphine dependence \cite{mamiya2001involvement}, and it can reduce heroin seeking relapse induced by cues or heroin initiation, by inhibiting heroin reward and seeking behavior \cite{lai2014phosphodiesterase}.

Baicalein is a flavonoid compound isolated from wood butterflies. It can effectively block morphine induced anti pain effects and delayed paw withdrawal, suggesting that it may regulate the pharmacological effects of opioid drugs by antagonizing opioid receptors \cite{singh2024baicalein}. Further research has found that baicalein, as an effective antagonist of $\mu$ - and  $\delta$ - opioid receptors, can reverse the cAMP inhibition mediated by agonists. This provides important scientific evidence for the development of therapeutic drugs related to opioid receptor activity regulation.

Ondelopran is a non selective opioid receptor antagonist that exhibits high affinity for the three classic opioid receptors ($\mu$, $\kappa$, and $\delta$)  \cite{fischler2022off}, providing potential new ideas for developing novel interventions for opioid addiction. 

To further evaluate the development potential of these three candidate drugs, we conducted a comprehensive ADMET property evaluation. As shown in Fig. \ref{FIG7_F2RL3_ADMET}, the evaluation results are presented in the form of a combination of radar images (upper part) and chemical structures (lower part), showing the ADMET properties of baicalein, ondelopran, and rolipram, respectively. The results indicate that the key ADMET indicators of these three drugs are within the ideal range, demonstrating good prospects for drug development.

\subsubsection{ADMET analysis} 

In the preliminary work of this study, in order to systematically evaluate the pharmacological properties of candidate drugs, we focused on examining the ADMET properties of six compounds, and the analysis results are presented in detail in Figs. \ref{FIG5_GPR84_ADMET} and \ref{FIG7_F2RL3_ADMET}, respectively.
As shown in Fig. \ref{FIG5_GPR84_ADMET}, the analysis results for epicatechin, lobeline, and axelopran indicate that the 13 key ADMET parameters (such as absorption, distribution, metabolism, excretion, and toxicity related indicators) of these three compounds are all within the ideal threshold. This result strongly demonstrates their good pharmacokinetic characteristics and low toxicity risk, laying a solid foundation for subsequent in vivo pharmacological studies.

Similarly, the ADMET properties of baicalein, ondelopran, and rolipram in Fig. \ref{FIG7_F2RL3_ADMET} also exhibit highly ideal characteristics. All its indicators meet the standards for optimizing lead compounds, further confirming that this group of molecules has significant advantages in oral bioavailability, metabolic stability, and safety. Overall, the ADMET evaluation results of these two compounds indicate that they have good development prospects.

Addtionally, we evaluated these six compounds based on criteria such as blood-brain barrier (BBB) permeability, P-gp inhibition, and P-gp substrate interactions, as shown in Table \ref{tab:compound_performance}. Epicatechin and baicalein excel in all aspects, demonstrating superior BBB penetration and minimal impact on P-gp. Lobeline and ondelopran show strong performance in P-gp but weaker BBB penetration, potentially limiting their application in central nervous system therapies. Axelopran exhibits good BBB penetration and P-gp inhibition but a low likelihood of being a P-gp substrate. Rolipram has poor BBB penetration, moderate P-gp inhibition, and is readily recognized by P-gp. These findings are crucial for guiding drug design and assessing clinical potential.

\begin{table}[htbp]
\centering
\caption{The performance of six candidate compounds evaluated in three important metrics, including BBB permeability and probability of P-gp inhibition and substrate interaction.}
\label{tab:compound_performance}
{\scriptsize
\begin{tabular}{|>{\centering\arraybackslash}m{2cm}|>{\centering\arraybackslash}m{2.5cm}|>{\centering\arraybackslash}m{2.5cm}|>{\centering\arraybackslash}m{4cm}|}
\hline
Compound & BBB Permeability & P-gp Inhibition & P-gp Substrate Interaction \\ \hline
Epicatechin & Excellent & Excellent & Excellent \\ \hline
Lobeline & Medium & Excellent & Excellent \\ \hline
Axelopran & Excellent & Excellent & Poor \\ \hline
Baicalein & Excellent & Excellent & Excellent \\ \hline
Onodelopran & Poor & Excellent & Excellent \\ \hline
Rolipram & Poor & Medium & Excellent \\ \hline
\end{tabular}
}
\end{table}

\section{Methods} \label{Methods}

\subsection{DEG analysis and PPI network}

The gene expression datasets used for DEG analysis in this study were all sourced from the public GEO database. We obtained 7 datasets from this database, including GSE174409, GSE182321, GSE194368, GSE210206, GSE210682, GSE260711, and GSE167922. In terms of DEG recognition, we used the Seurat software package for processing the GSE260711 dataset with 10x Genomics data format, and the values of min.pct parameter and adjust-p-value are set to 0.25 and 0.01, respectively. For the other six datasets with counting matrix, DESeq2 package was used for analysis, the settings for Log2FoldChange and p-value are shown in Table \ref{tab:DEG_parameters}. Additionally, the PPI network information between DEGs is sourced from the STRING database.

\begin{table}[htbp]
\centering
\caption{The summary of Log2FoldChange and p-value settings of DESeq2 package used in DEG analysis on six datasets.}
\label{tab:DEG_parameters}
{\scriptsize
\begin{tabular}{|>{\centering\arraybackslash}m{1.5cm}|>{\centering\arraybackslash}m{2.5cm}|>{\centering\arraybackslash}m{2cm}|}
\hline
Dataset & Log2FoldChange & p-value \\ \hline
GSE174409 & 1 & 0.01 \\ \hline
GSE182321 & 1 & 0.05 \\ \hline
GSE194368 & 1 & 0.01 \\ \hline
GSE210206 & 1 & 0.01 \\ \hline
GSE210682 & 3 & 0.01 \\ \hline
GSE167922 & 1 & 0.05 \\ \hline
\end{tabular}
}
\end{table}

\subsection{Multiscale topological differentiation of network}

PPI network plays a crucial role in cellular function and disease expression. Therefore, analyzing the PPI network is of great significance for the understanding of the complex physiological mechanisms within living organisms. In the present study, we employed the multiscale topological differentiation method proposed by Du et al. to analyze the PPI network \cite{du2024multiscale}. This method combines the principles of PST and PH, and its core lies in analyzing the differences in topological invariants and geometric features of PPI networks before and after node deletion. Through this process, the aim is to identify genes that occupy key positions in the network structure.

\subsubsection{Persistent homology}

For a $k$-simplex $\sigma_i^k $, its $k$-chain $[\sigma^k]$ is a linear combination of $k$-simplexes $\sum_{i}^k \alpha_i \sigma_i^k$, where $\alpha_i $ is a coefficient. The collection of all $k$-chains of the simplicial complex $K$ together with addition forms an Abelian group $C_k(K, \mathbb{Z}_2)$.
The homology of the topological space is represented by a series of Abelian groups.

For a given $k$-simplex $\sigma^k = \{v_0, v_1, v_2, \cdots, v_k\}$, its boundary can be represented as:
\begin{equation}
\partial_k \sigma^k = \sum_{i=0}^{k} \{v_0, v_1, \ldots, \hat{v}_i, \ldots, v_k\},  
\end{equation}
where $\partial_k$ is the boundary operator, defined as $\partial_k : C_k \rightarrow C_{k-1}$, $C_k$ is an Abelian group.

Using the boundary operator, cyclic groups and boundary groups can be defined. The $k$-th cycle group $Z_k$ and the $k$-th boundary group $B_k$ are subgroups of $C_k$, and can be defined as:

\begin{equation}
Z_k = \text{Ker} \partial_k = \{c \in C_k \mid \partial_k c = \varnothing\}, 
\end{equation}

\begin{equation}
B_k = \text{Im} \partial_{k+1} = \{c \in C_k \mid \exists d \in C_{k+1}, c = \partial_{k+1} d\}. 
\end{equation}

With all the above definitions, we can introduce the concept of homology. Specifically, the  $k$-th homology group $H_k$ is generated by the $k$-th cycle group  $Z_k$ and the $k$-th boundary group  $B_k: H_k = Z_k / B_k$.  If two $k$-th cycle elements are equivalent modulo the $k$-th boundary elements, they are said to be homologous. From the fundamental theorem of $k$-generated Abelian groups, it is known that the $k$-th homology group  $H_k$ can be expressed as a direct sum:

\begin{equation}
H_k = Z \oplus \cdots \oplus Z \oplus Z_{p_1} \oplus \cdots \oplus Z_{p_n} = Z^{\beta_k} \oplus Z_{p_1} \oplus \cdots \oplus Z_{p_n}, 
\end{equation}
in which $\beta_k$, the rank of the free subgroup, is the $k$-th Betti number. $Z_{p_i}$ is the  torsion subgroup with torsion coefficients $(p_i = 1, 2, \ldots, p_n)$, the power of prime number. Therefore, regardless of when $H_k$ is torsion free. The Betti number can be simply calculated as:

\begin{equation}
\beta_k = \text{rank} H_k = \text{rank} Z_k - \text{rank} B_k.
\end{equation}

In topology, the cycle elements in  $H_k$ form a $k$-dimensional cycle or void, which is not derived from higher-dimensional boundary elements. The geometric significance of Betti numbers is as follows: $\beta_0$ represents the number of isolated components,  $\beta_1$ is the number of one-dimensional cycles or loops,  $\beta_2$ describes the number of two-dimensional voids or holes, and the Betti sequence $\{\beta_0, \beta_1, \beta_2, \ldots\}$ describes the intrinsic topological properties of the system.

For a simplicial complex $K$, the filtration is defined as a sequence of subcomplexes:

\begin{equation}
0 = K^0 \subset K^1 \subset \cdots \subset K^m = K. 
\end{equation}

Generally speaking, the abstract simplicial complexes generated by the filtration provide a multi-scale representation of the corresponding topological space, from which the homology groups can be evaluated to reveal topological features. On this basis, the concept of persistence is introduced to describe the persistent topological features. Thus, the $p$-persistent $k$-th homology group $H_k^{t,p}$ is

\begin{equation}
H_k^{t,p} = Z_k^t / (B_k^{t+p} \cap Z_k^t).
\end{equation}

By investigating the persistence patterns of these topological features, the so-called PH can capture the intrinsic properties of the underlying space from a point cloud.

\subsubsection{Persistent spectral graph theory}

PPI network is commonly modeled as a graph in which nodes represent proteins and edges represent pairwise interactions. Yet the classical graph-theoretic framework has an inherent limitation: it cannot adequately capture higher-order interactions. To overcome this challenge, Du et al. introduced  the topological approach based on persistent spectral graph theory, opening a new perspective on PPI networks \cite{du2024multiscale}. Persistent spectral graph theory encodes relationships that transcend simple pairwise contacts, retains far richer "shape" information of the network, and enables the analysis of high-dimensional topological structures\cite{wang2020persistent}. Building on this foundation, persistent spectral graph theory deepens our understanding of these complex architectures. The theory tracks how topological and geometric features evolve across a continuous parameter, thereby uncovering hidden multiscale organizational patterns embedded in PPI data. Its core mechanism is the analysis of a filtration, a nested sequence of simplicial complexes parameterized by a real variable. Mathematically, a simplicial complex is assembled from a finite set of simplices that constitute its structure.

A $k$-simplex $\sigma^k$ is composed of the convex hull of $k+1$ affinely independent points $v_0, v_1, v_2, \cdots, v_k$ :

\begin{equation}
\sigma^k := [v_0, v_1, v_2, \cdots, v_k] = \left\{ \sum_{i=0}^{k} \lambda_i v_i \ \middle| \ \sum_{i=0}^{k} \lambda_i = 1; \lambda_i \in [0,1], \forall i \right\}.
\end{equation}

A filtration refers to a sequence of simplicial complexes $\{K_t\}_{t \in \mathbb{R}^+}$ defined by a parameter $t$, and the simplicial complexes within the sequence are ordered by inclusion. The filtration process has the following properties:

1. For two parameter values $t' < t''$, we have $K_{t'} \subseteq K_{t''}$;

2. In the filtration, there are only a finite number of shape changes, and we can find at most $n$ filtration parameters such that: 
\begin{equation}
\varnothing \subsetneq K_{t_1} \subsetneq K_{t_2} \subsetneq \cdots \subsetneq K_{t_n} = K,
\end{equation}
where $K$ is the largest simplicial complex.

Let $t_i$ be the smallest filtration parameter at which an $i$-th shape change is observed. Then, for any filtration parameter $t$, the corresponding simplicial complex can be determined by equation (3):

\begin{equation}
K_t = 
\begin{cases} 
K_{t_i}, & \text{if } t \in [t_{i-1}, t_i), i \leq n, \\
K_{t_n}, & \text{if } t \in [t_n, \infty).
\end{cases}
\end{equation}

For each subcomplex $K_t$ in the filtration, a series of chain complexes can be constructed:

\begin{equation}
\left\{
\begin{array}{c}
\cdots \xrightleftharpoons[\partial_{k+2}^{r_1}]{\partial_{k+2}^t} C_{k+1}^t \xrightleftharpoons[\partial_{k+1}^{r_1}]{\partial_{k+1}^t} C_k^t \xrightleftharpoons[\partial_k^{r_1}]{\partial_k^t} \cdots \xrightleftharpoons[\partial_1^{r_1}]{\partial_1^t} C_0^t \xrightleftharpoons[]{}\begin{array}{c}
\partial_0^t \\
\partial_0^{t^*} 
\end{array}
\varnothing
\end{array}
\right\},
\end{equation}
in which $C_k^t = C_k(K_t)$ is its chain group, $\partial_k^t : C_k(K_t) \rightarrow C_{k-1}(K_t)$ is its $k$-th boundary operator, ${\partial_k^t}^*$is the adjoint operator of the boundary operator $\partial_k^t$, and it is relative to an inner product defined on a chain group.

For a $p$-persistent chain group $\mathbb{C}
_k^{t,p} \subseteq C_k^{t+p}$ on the boundary in $C_{k-1}^t$, it is defined as

\begin{equation}
\mathbb{C}
_k^{t,p} = \left\{ \alpha \in C_k^{t+p} \mid \partial_k^{t+p}(\alpha) \in C_{k-1}^t \right\},
\end{equation}
here  $\partial_k^{t+p} : C_k^{t+p} \rightarrow C_{k-1}^{t+p}$ is the $k$-th boundary operator of the chain group $ C_k^{t+p}$.

Then, define a $p$-persistent boundary operator $\eth_k^{t,p}$ as the restriction of $\partial_k^{t+p}$ on the $p$-persistent chain group $ \mathbb{C}_k^{t,p}$ :

\begin{equation}
\eth_k^{t,p} = \partial_k^{t+p} \big|_{\mathbb{C}_k^{t,p}} : \mathbb{C}_k^{t,p} \rightarrow C_{k-1}^t. 
\end{equation}

The $p$-persistent $k$-th combinatorial Laplacian operator $\Delta_k^{t,p} : C_k(K_t) \rightarrow C_k(K_t) $ is given by :

\begin{equation}
\Delta_k^{t,p} = \eth_{k+1}^{t,p} (\eth_{k+1}^{t,p})^* + (\partial_k^t)^* \partial_k^t.
\end{equation}

The matrix representations of the boundary operators $\eth_{k+1}^{t,p}$  and $\partial_k^t$ are denoted as $\mathcal{B}_{k+1}^{t,p} \text{ and } \mathcal{B}_k^t$, respectively.

$\Delta_k^{t,p}$ 's Laplacian matrix is:
\begin{equation}
\mathcal{L}_k^{t,p} = \mathcal{B}_{k+1}^{t,p} (\mathcal{B}_{k+1}^{t,p})^\top + (\mathcal{B}_k^t)^\top \mathcal{B}_k^t. 
\end{equation}

Since the Laplacian matrix $\mathcal{L}_k^{t,p}$ is symmetric and positive semi-definite, all its eigenvalues are non-negative real numbers:

\begin{equation}
S_k^{t,p} = \text{Spectra}(\mathcal{L}_k^{t,p}) = \{(\lambda_1)_k^{t,p}, \{(\lambda_2)_k^{t,p}, \cdots, (\lambda_N)_k^{t,p}\},
\end{equation}
here $N$ represents the dimension of the standard basis for $C_k^t$, which is  the number of $k$-simplices, and $\mathcal{L}_k^{t,p}$  has dimension  $N \times N$. The $k$-th persistent Betti number $\beta_k^{t,p}$ can be obtained from the eigenvalues of $\mathcal{L}_k^{t,p}$:

\begin{equation}
\beta_k^{t,p} = \text{dim}(\mathcal{L}_k^{t,p}) - \text{rank}(\mathcal{L}_k^{t,p}) = \text{null}(\mathcal{L}_k^{t,p}) = \#\{i \mid(\lambda_i)_k^{t,p} \in S_k^{t,p}, \text{ and } (\lambda_i)_k^{t,p} = 0\}.
\end{equation}

PST provides insights at the geometric level, surpassing purely topological persistence analysis, and can extract information from the Laplacian operators of spectral persistence complexes. Specifically, persistent Betti numbers offer information about the constancy of topological structures, while transformations in geometric shapes can be distinguished through the non-harmonic components of the spectra.

\subsubsection{Key gene identification via network topological differentiation}

In the method of Du et al., they used PST and PH to evaluate the importance of a single gene in a PPI network represented as $G = (V, E)$. $G_m $ means the subgraph obtained by removing the $m$-th vertex $v_m$ and all edges connected to $v_m$. For a given threshold $T$, the distance between two nodes $v_i$ and $v_j$ in $G$ is defined as: 

\begin{equation}
D_{ij} = 
\begin{cases} 
1 - s_{ij}, & \text{if } s_{ij} > T, \\
\infty, & \text{otherwise.}
\end{cases} 
\end{equation}
here, $s_{ij}$ represents the combined score of the interactions between proteins $v_i$ and $v_j$ in the STRING database.

For each network, ten filtration parameters are uniformly selected from the interval $(0,1-T)$. For each Laplacian spectrum, the harmonic spectra count and the five statistical descriptors of the non-harmonic spectra (minimum, maximum, mean, standard deviation, and sum) are calculated. For each network, its attributes are encapsulated into a vector $f_G = \Theta(G)$, and then the feature vector  $f_{G_m} = \Theta(G_m)$ of the subgraph $G_m$ after perturbation at vertex $v_m$ is obtained. Finally, the importance of vertex $v_m$ is quantified by calculating the Euclidean distance between the feature vectors $f_G$ and $f_{G_m}$:

\begin{equation}
S_m^G = \text{distance}(f_G, f_{G_m}).
\end{equation}

This measure reflects how the removal of gene $v_m$ affects the structure of the network in terms of topology and geometric shape, thereby reflecting the impact of gene $v_m$ within the PPI network $G$.

\subsection{Machine learning-based drug repurposing}\
\subsubsection{Data preparation}

In order to train the ML model we constructed, we obtained the inhibitor dataset of key genes from the CHEMBL database. These datasets include SMILE strings and biological activity labels ($\mathrm{IC_{50}}$ and $K_i$) for molecular compounds, where $\mathrm{IC_{50}}$ (half maximal inhibitory concentration) refers to the concentration of inhibitors (such as drugs, antibodies, etc.) required to inhibit a certain biological activity (such as enzyme activity, cell proliferation, viral infection, etc.) by 50\% under specific experimental conditions. The $K_i$ value (inhibition constant) is the dissociation constant of a complex formed by the binding of an inhibitor to an enzyme in an enzyme inhibition reaction, reflecting the affinity between the inhibitor and the enzyme. The smaller the $K_i$ value, the higher the affinity between the inhibitor and the enzyme. To adjust these labels to fit the BA of our model, we use the relationship $K_i = \mathrm{IC_{50}}/2$ to approximately convert $\mathrm{IC_{50}}$ to $K_i$, and then use these labels to calculate BA values based on the formula: $BA = 1.3633 \times \log_{10}(K_i) \ (\text{kcal/mol})$ \cite{kalliokoski2013comparability}. If a single molecule has multiple labels, we calculate the average of these labels as its final affinity value. Additionally, we only studied small molecule drugs classified as "approved" and "investigational" retrieved from the DrugBank (Version 5.1.13) database.

\subsubsection{Molecular fingerprints}

Molecular fingerprinting refers to the technology of converting chemical molecules into digital codes and is a common form of representation. In order to characterize the molecular structure, this study selected three different fingerprint generation strategies. Two of these strategies integrate the natural language processing (NLP) techniques: one utilizes a model based on bidirectional transformers \cite{chen2021extracting}, and the other employs a sequence to sequence autoencoder architecture \cite{winter2019learning}. These two NLP inspired methods both utilize pre-trained models to encode standardized SMILES strings into a 512 dimensional latent vector space. Additionally, to enrich the diversity of methods, this study also adopted a classic topological fingerprint, which uses the RDKit toolkit to calculate and generate two-dimensional extended connectivity fingerprints (ECFPs) \cite{rogers2010extended} as molecular descriptors.

\paragraph{Bidirectional encoder transformer fingerprint (BET-TP)}

Chen et al. proposed a self-supervised learning platform (SSLP) in 2021\cite{chen2021extracting}, which aims to extract predictive representations from unlabeled molecular data. This platform consists of four core modules: pre-training dataset module, dataset analysis module, pre-training model module, and fine-tuning module. Its pre-training dataset integrates three publicly available chemical databases: ChEMBL, PubChem, and ZINC. In the pre-training model module, the platform adopts BET based on Transformer for self-supervised learning. Specifically, this module achieves the learning process of the model by randomly masking some symbols in the SMILES string and training the model to predict these masked symbols. The dataset analysis module uses Wasserstein distance to quantify the similarity between different datasets, and uses ridge regression model to select the most suitable pre-training model for specific tasks. The fine-tuning module is responsible for further optimizing the selected pre-trained model based on a specific dataset, in order to generate molecular fingerprints highly relevant to the task.

The BET model used on this platform has an attention mechanism as its core mechanism, which can effectively capture the importance of each symbol in the input sequence. Moreover, the design of independent position embedding significantly enhances the parallel processing capability of the transformer model. The basic architecture of the BET model is consistent with traditional transformer encoders, consisting of eight encoder layers. Each encoder layer consists of a self-attention layer and a fully connected feedforward network, and each encoder layer is configured with eight self-attention heads.

Before inputting SMILES strings into the BET model for training, a series of preprocessing operations are required to ensure consistency in data format and make it suitable for model training. The preprocessing steps include: dividing the SMILES string into 51 basic symbol units including element symbols, parentheses, etc., and adding "<s>" and "</s>" at the beginning and end of each SMILES string, respectively, to indicate the boundary of the sequence. These steps uniformly limit the length of the input sequence to no more than 256 characters. For SMILES strings with a length of less than 256 characters, the "<pad>" symbol is used to fill in. The core of self- supervised learning lies in using unlabeled data to train models by constructing data mask pairs. The specific implementation of masking operation is to randomly select 15\% of the SMILES string for masking processing, where 80\% of the symbols are replaced with mask markers, 10\% of the symbols remain unchanged, and the remaining 10\% of the symbols are randomly replaced with other symbols. The model receives the masked SMILES string as input and learns to predict the original symbol of the masked position based on it. Through this process, the model is able to learn and understand the intrinsic connections between symbols in SMILES strings during the pre-training stage, and then infer the masked symbols, ultimately achieving effective understanding of SMILES language.

In this article, we use SMILES strings obtained from the CHEMBL database as input to generate BET molecular fingerprints.

\paragraph{Sequence-to-sequence auto-encoder fingerprint (AE-TP)}
Wente et al. proposed a deep learning based approach in 2019 aimed at generating continuous and data-driven molecular descriptors \cite{winter2019learning}. The core architecture of this method draws on the neural machine translation (NMT) model, which includes two key components: an encoder and a decoder. This model learns a low dimensional continuous vector representation by mapping a certain representation of a molecule (such as a SMILES string) to another representation that is semantically equivalent but has a different syntactic structure. This vector representation can be used as a molecular descriptor for subsequent chemical informatics analysis tasks.

The encoder is responsible for compressing and encoding the input SMILES string into a low dimensional continuous vector, namely latent representation. In terms of specific implementation, the encoder first receives the SMILES string that has been tokenized and uniquely encoded, and then uses a series of neural network layers (such as convolutional neural network or recurrent neural network ) to extract key information about the molecular structure, ultimately condensing it into a fixed dimensional vector. This vector theoretically contains all the necessary information about the molecular structure, providing a foundation for the subsequent decoding and translation process.

The decoder is responsible for reconstructing the target sequence (such as standard SMILES or InChI) from the low dimensional continuous vectors output by the encoder. It receives potential representations from the encoder and gradually generates the target sequence through a series of neural network layers. The decoder outputs a probability distribution of a character at each step, and the final target sequence is constructed by selecting the character sequence with the highest probability. In order to improve the stability of the training phase and the accuracy and efficiency of the reasoning phase, the study also adopted teacher forcing techniques for training.

The training process of the model is guided by minimizing the translation error between the input sequence and the target sequence. Specifically, the optimization of model parameters is achieved by minimizing the cross entropy loss function, which measures the difference between the predicted distribution of the model and the true target sequence. Additionally, to further enhance the model's understanding of the intrinsic properties of molecular structures, we also introduce an auxiliary classification task that requires the model to predict specific molecular properties (such as physical and chemical properties). Through this multi-task learning approach, the model not only optimizes the translation accuracy of molecular representations, but also deepens the learning of key features of molecular structures, thereby generating more informative and explanatory molecular descriptors.

\paragraph{Extended-connectivity fingerprints (ECFP)}

ECFP \cite{rogers2010extended} is a topological fingerprinting method proposed by David Rogers in 2010 for molecular characterization, which is particularly suitable for modeling structure-activity relationships.

The generation process of ECFP is based on an improved version of the Morgan algorithm. The Morgan algorithm was originally designed to solve the problem of molecular isomorphism, which involves identifying molecules with different atomic numbers but essentially the same structure. The ECFP algorithm has made key modifications to the Morgan algorithm: it sets a predetermined upper limit on the number of iterations and retains the intermediate generated atomic identifiers at the end of each iteration, rather than pursuing absolute uniqueness like the Morgan algorithm. The generation process of ECFP mainly includes three stages: initial atomic identifier allocation, iterative update of identifiers, and removal of duplicate identifiers.

In the initial atomic identifier allocation stage, the system assigns an initial integer identifier to each atom in the molecule. After entering the iteration update identifier stage, the identifier of each atom will be dynamically updated based on the current identifier of its neighboring atoms, and this process will be repeated within a preset number of iterations. Finally, in the stage of removing duplicate identifiers, the algorithm identifies and removes different identifiers generated during the iteration process that actually represent the same substructure, ensuring the uniqueness and accuracy of the fingerprint.

In this study, we utilized the RDKit chemical informatics toolkit to generate ECFP features. The specific parameters are set to a fingerprint generation radius of 2, and the length (i.e. number of bits) of the fingerprint vector is set to 2048.

\subsubsection{Machine learning models}
SVM is a powerful algorithm developed within the framework of supervised learning, widely used to solve classification and regression problems. The core mechanism is to explore an optimized hyperplane from the feature space to clearly distinguish data points of different categories, thereby achieving classification or regression prediction. Specifically, SVM can use non-linear kernel functions to map raw data to a higher dimensional feature space. In this space, the algorithm searches for hyperplanes that can effectively separate data and adjusts their positions to maximize the classification interval - that is, the distance between the hyperplane and the nearest data points from different categories (i.e. support vectors). The process of maximizing this interval helps to improve the generalization ability of the model, ensuring that the decision boundary is as far away from the data points of each category as possible, and the resulting hyperplane can be represented as a linear combination of these key support vectors \cite{azlim2023comparative}.

RF is a classic ensemble learning technique. Its operating mechanism lies in constructing and combining multiple decision tree models during the training phase. For classification tasks, the final prediction results are generated by voting on these trees; For regression tasks, the average of all tree predictions is taken as the output. The uniqueness of this method lies in that each decision tree is constructed based on a random subset of the original training data, and at each split node of the tree, only the optimal partition criterion is selected from a random subset of all features. This introduced dual randomness (sample randomness and feature randomness) promotes the generation of diverse tree models. When these diverse individual trees are combined through ensemble, they can form a more superior overall model, effectively improving the model's generalization ability and significantly reducing the risk of overfitting \cite{ahmed2025deep}.

GBDT model was first proposed by Jerome Friedman in 2001 \cite{friedman2001greedy} and has now become a highly influential ensemble learning paradigm in the fields of ML and data mining, suitable for regression and classification tasks. The basic principle is to iteratively combine multiple learners with relatively weak performance (individual decision trees), gradually optimize the model performance, and ultimately construct a strong learner with significantly enhanced predictive ability. Compared to other ML methods, GBDT typically exhibits better performance, especially when dealing with small sample datasets \cite{dou2023machine}. The advantage of this algorithm lies in its resistance to overfitting and insensitivity to hyperparameters. This enables GBDT to achieve satisfactory prediction results in various practical application scenarios, such as the recognition of anti-cancer peptides \cite{li2023acp} and the prediction of disease-related genes \cite{helmy2022predicting}.

In the implementation, in order to reduce the potential impact of random factors during model training, we trained each model with different random seeds 10 times and used the average of the predicted results obtained from these 10 independent runs as the final output of the model.

\section{Conclusion} \label{Conclusion}

The abuse and addiction of opioid drugs constitute an increasingly severe global health crisis. However, the current clinical available treatment strategies are limited and have varying effects. Therefore, developing new and effective intervention treatment has become an urgent task. To address this challenge, the present work proposes a systematic drug reuse strategy integrating bioinformatics and mathematical tools with machine learning methods, aimed at identifying potential repurposed drug candicates for opioid addiction.

We first collected and did the meta-analysis of seven publicly  transcriptomic datasets related to opioid addiction. By applying a novel multi-scale topological differentiation analysis algorithm, we accurately identified a group of key genes that play a vital role in the opioid addiction process. After pathway validation and functional enrichment analysis with existing scientific literature, we constructed a core gene set related to opioid addiction consisting of 1865 genes. This gene set has laid a molecular foundation for subsequent target identification and drug screening.

Based on the protein targets encoded by the key genes mentioned above, we identified 72 inhibitor datasets with 46977 compounds from the CHEMBL database as training data for downstream machine learning prediction. In terms of molecular representation, we adopted a feature fusion strategy that aggregated three different fingerprints together, including transformer-based fingerprint (BET-TP), auto encoder-based fingerprint (AE-TP), and a classical two-dimensional fingerprint ECFP. Subsquently, we built machine learning models by integrating three different regressors with fused fingerprint, including SVM, RF, and GBDT. These models were used for large-scale virtual screening of over 6000 approved and investigational drugs in the DrugBank database, predicting their binding affinity to core targets. We found that SVM model had best performance in the prediction with smallest RMSE values. Additionally, the resulting promising drug candicates were analyzed by molecular docking studies and further screened for their ADMET properties to preliminarily assess their in vivo pharmacokinetic characteristics and safety. However, the identified promising drugs require additional in vivo validation to determine their safety and effectiveness in reducing substance addiction.

In summary, this study not only successfully identified a series of drug candidates with repurposing potential for opioid addiction, but more importantly, we established a powerful computational framework that connects the transcriptome data with drug repurposing study. This framework provides a new perspective for understanding the pathological mechanisms of complex diseases and accelerating the process of drug discovery. Its application scope can be expanded across a spectrum of diseases and transcriptomic datasets. 

\section*{Acknowledgments}
This work was supported in part by MSU Research Foundation. The work of Huahai Qiu and Bengong Zhang was supported by the National Natural Science Foundation of China under Grant No.12271416 and No.12371500, respectively. 

\section*{Data Availability}
The data and source code of this study are freely available at GitHub (https://github.com/hahaha3758/Drug\_repurposing).

\section*{Conflict of interest declaration}
We declare we have no competing interests.

\bibliographystyle{abbrv}
\bibliography{ref}
\end{document}